\documentclass[11pt,twocolumn]{article}
\usepackage[utf8]{inputenc}
\usepackage[T1]{fontenc}
\usepackage{microtype,graphicx,subcaption,booktabs,float}
\usepackage{tikz,tikz-feynman}
\usepackage[dvipsnames]{xcolor}
\usepackage{url,hyperref}
\usepackage{authblk}
\usepackage{stfloats} % [b] positioning in figure*
\hypersetup{
    colorlinks = true,
    linkcolor  = BrickRed,
    citecolor  = teal, 
    linkbordercolor = {0 1 0},
    filecolor = magenta,
    urlcolor = BrickRed,
    linktoc=all,
}
\usepackage[style=numeric,sorting=none,doi=false,url=false,isbn=false,eprint=false]{biblatex}
\renewbibmacro{in:}{}
\usepackage{titlesec}

\titleformat{\section}
  {\centering\normalsize\bfseries}
  {\thesection}
  {0.5em}
  {\MakeUppercase}

\titleformat{\subsection}
  {\centering\small\bfseries}
  {\Alph{subsection}.}
  {0.5em}
  {\MakeUppercase}

\usepackage{geometry}
\usepackage{mathtools,amssymb,amsthm,bm,cancel}
\usepackage{mathrsfs}   % For \lden, \eden font
\newcommand{\lden}{\mathscr{L}}
\newcommand{\eden}{\mathscr{E}}

\usepackage{cleveref}
\usepackage{slashed}         % Feynman slash notation \slashed{p}
\usepackage{siunitx}
\usepackage{physics}
\AtBeginDocument{\RenewCommandCopy\qty\SI}

\renewcommand{\thesection}{\Roman{section}.}

\titleformat{\subsubsection}
  {\small\bfseries}
  {}
  {0pt}
  {}
\usepackage{tocloft}

\makeatletter
\renewcommand{\numberline}[1]{\hb@xt@\@tempdima{\hfil#1}\hspace{0.5em}}
\makeatother

\AddToHook{bfseries}{\boldmath}

\begin{document}
% Rename table and figure names
\renewcommand{\tablename}{Table}
\renewcommand{\figurename}{Fig.}

% Faster greek alphabet
\newcommand{\al}{\alpha}
\newcommand{\be}{\beta}
\newcommand{\ga}{\gamma}
\newcommand{\de}{\delta}
\newcommand{\ep}{\epsilon}
\newcommand{\sig}{\sigma}
\newcommand{\ome}{\omega}

% Quicker \mu & \nu partials
\newcommand{\p}{\partial}
\newcommand{\pmu}{\partial_{\mu}}
\newcommand{\pnu}{\partial_{\nu}}
\newcommand{\upmu}{\partial^{\mu}}
\newcommand{\upnu}{\partial^{\nu}}

\newcommand{\pal}{\partial_{\alpha}}
\newcommand{\pbe}{\partial_{\beta}}
\newcommand{\pga}{\partial_{\gamma}}
\newcommand{\upal}{\partial^{\alpha}}
\newcommand{\upbe}{\partial^{\beta}}
\newcommand{\upga}{\partial^{\gamma}}

\newcommand{\pj}{\partial_j}
\newcommand{\pk}{\partial_k}
\newcommand{\mybar}[2]{\left. #1 \right|_{#2}}

% Quantum mechanics
\newcommand{\dg}{^{\dagger}}

%Toms commands
\definecolor{TW-color}{RGB}{100,0,100}
\definecolor{Error-color}{RGB}{250,0,0}
\newcommand{\TWedit}[1]{{\color{TW-color}#1}}
\newcommand{\Error}[1]{{\color{Error-color}#1}}

\definecolor{AA-color}{RGB}{0,47,167}
\newcommand{\AAedit}[2]{%
  \fcolorbox{AA-color}{AA-color!10}{\textcolor{AA-color}{\sffamily\bfseries\tiny #1}}%
  \,\textcolor{AA-color}{#2}%
}

\renewcommand{\thefootnote}{\alph{footnote}}

% Title and abstract 

\title{\bfseries Nuclear geometry under external magnetic field \\ in the Skyrme-Maxwell model}

\author[1,a]{Christoph Adam}
\author[1,b]{Andrés R. Anadón}
\author[2,c]{Carlos Naya}
\author[3,d]{Thomas Winyard}

\affil[1]{Departamento de Física de Partículas, Universidad de Santiago de Compostela and Instituto Galego de Física de Altas Enerxías (IGFAE), E-15782 Santiago de Compostela, Spain}
\affil[2]{Universidad de Alcalá, Departamento de Física y Matemáticas, 28805 Alcalá de Henares, Spain}
\affil[3]{Division of Mathematics, University of Dundee, Dundee DD1 4HN, United Kingdom}

\date{\vspace{-3.0ex}(Dated: \today)}

\twocolumn[
\maketitle
\begin{abstract}
We find static solutions of the $U(1)$ gauged Skyrme-Maxwell model in a constant external magnetic field, for baryon numbers $B\leq 6$. The field acts as an isospin-breaking source, stretching the skyrmion along its direction, while its size grows and then shrinks as the induction increases. We obtain the energy minimizers with a GPU implementation of the arrested Newton flow (ANF) algorithm. Gauged skyrmions have less energy and are more strongly bound than their ungauged counterpart. At weak magnetic field, the solutions are small deformations of the ungauged skyrmions, including regions of negative baryon density in sectors $B=3,5,6$. For intermediate field, novel geometries appear, notably a $(B\!-\!1)$-toroidal chain. High field values lead to axially symmetric toroidal structures for all $B$. 
\end{abstract}
\vspace{1em}
]

\footnotetext[1]{\texttt{christoph.adam@usc.es}}
\footnotetext[2]{\texttt{andresrodriguez.anadon@usc.es}}
\footnotetext[3]{\texttt{carlos.naya@uah.es}}
\footnotetext[4]{\texttt{twinyard001@dundee.ac.uk}}
\setcounter{footnote}{0}

\section{Introduction}\label{sec:intro}

Our modern understanding of nuclear and subnuclear systems rests on Quantum Chromodynamics (QCD), the fundamental theory of the strong interaction between quarks and gluons. The strength of this force depends heavily on the energy scale. At high energies ($\gtrsim\SI{1}{\giga\electronvolt}$), the coupling constant becomes small and the quarks and gluons behave similarly to free particles, a property known as asymptotic freedom. In this regime QCD can be treated perturbatively, much like Quantum Electrodynamics (QED). Conversely, at low energy scales ($\lesssim\SI{200}{\mega\electronvolt}$), the strong force tightly binds quarks and gluons into colourless hadrons. The process of confinement remains one of the central open problems in theoretical physics. In this non-perturbative regime, QCD becomes analytically intractable, motivating the search for alternative effective field theories (EFTs).

A foundational proposal for an EFT of nuclear matter was introduced by Tony Skyrme \cite{Skyrme1961,Skyrme1962}, in which baryons emerge as topological soliton solutions of a nonlinear meson theory. It was later shown \cite{Hooft1974,Witten1979} that QCD reduces to an effective theory of mesons in the large $N_c$ limit. In the Skyrme model, the baryon charge arises as a topological index, rather than as a Noether charge as in other EFTs, such as Chiral Perturbation Theory (ChPT) or Relativistic Mean-Field (RMF) models. Furthermore, it provides a unified framework to simultaneously describe mesons, baryons, and nuclei with very few parameters, although the quantitative predictions of the original model deviate from experimental values by roughly 20-30\% \cite{Adkins1983, Barnes1997}. To address these limitations, numerous refinements have been developed over recent decades including higher order derivative terms, coupling to mesons or quantizing the model. These inclusions yield more realistic nuclear binding energies, reproduce the linear growth of nuclear masses with the baryon number, capture the clustering structure of light nuclei and correctly order the excitation spectra of many complex systems such as carbon-12 and oxygen-16 \cite{Adam2010,Adam2010_2,Lau2014,Gillard2015,Gudnason2016,Halcrow2017,Naya2018}.

The Skyrme model can also be improved by gauging a subgroup of its global $SU(2)_L\times SU(2)_R$ symmetry, which adds interactions absent from the original theory. The $U(1)$ gauged model was first introduced to construct a semiclassical description of the Rubakov–Callan effect in nucleon decay \cite{CallanWitten1984}. That work also includes the QCD anomalies via the Wess-Zumino-Witten (WZW) effective action term. Even without this term, Piette and Tchrakian \cite{Piette2000} showed that the Skyrme-Maxwell model alone admits stable classical soliton solutions with a self-contained electromagnetic field, allowing the skyrmion to carry electric charge. In addition, an $SU(2)_L$ model has been used to study weak interactions in the Skyrme model \cite{d1984decoupling}, and $SU(2)$ models have also been shown to admit static solutions \cite{arthur1996so,brihaye2001monopoles}. Finally, a model derived from $SU(2)$ Yang–Mills theory, with the gauge symmetry broken to $U(1)$, was proposed in \cite{cork2022model} as a more physically realistic version of the Skyrme-Maxwell model, leading to lower binding energies.

In this paper we study the $U(1)$ gauged Skyrme-Maxwell model in the presence of an external magnetic field. Strong magnetic fields are found in a variety of physical systems. For instance, magnetic fields at the surface of magnetars can reach values of \SIrange{e10}{e11}{\tesla}, while observations of giant flares suggest that even stronger fields may exist in their interiors \cite{Duncan1992, Turolla2015}. In addition, heavy-ion collisions can generate magnetic fields as strong as \SI{e14}{\tesla} \cite{Skokov2009}. In our study, we find that skyrmions already undergo appreciable deformations at magnetic field strengths of the order of \SIrange{e14}{e15}{\tesla}. Our results show that external magnetic fields have a significant impact on both the geometry and the energy of gauged skyrmions, thereby influencing the structure of the corresponding nuclei. Throughout this work, we use the metric signature $(+,-,-,-)$.

This paper is organized as follows, \Cref{sec:model} introduces the $U(1)$ gauged Skyrme-Maxwell model and discusses the properties of its static minimisers. \Cref{sec:method} describes the numerical implementation of the arrested Newton flow (ANF) algorithm used to compute static skyrmion solutions. \Cref{sec:results} presents the numerical results for gauged skyrmions with $B \leq 6$. Finally, \Cref{sec:conclusions} summarises our results and discusses their applications.

\section{The Skyrme-Maxwell model}\label{sec:model}

The original Skyrme Lagrangian density is given by,
\begin{align}
\lden &= \frac{f_\pi^2}{16\hbar c}\Tr\left(\pmu U\dg \upmu U\right) \nonumber \\[4pt]
      &\quad + \frac{\hbar c}{32a_0^2}\Tr\Big[\big(\partial_{[\mu} U\dg \partial_{\nu]} U\big)
               \big(\partial^{[\mu} U\dg \partial^{\nu]} U\big)\Big] \label{lag:0} \\[4pt]
      &\quad - \frac{m_\pi^2 c^4 f_\pi^2}{8(\hbar c)^3}\Tr(\sig_0 - U), \nonumber
\end{align}
where the classical Skyrme field is $U\in SU(2)$ and we use $(\sig_1,\sig_2,\sig_3)$ to denote the 2-dimensional Pauli matrices, while $\sig_0 = 1$ is the identity matrix. The model depends on three free parameters: the pion mass $m_\pi$, the pion decay constant $f_\pi$ and $a_0$ which determines the spatial scale. We realise the $U(1)$ gauged Skyrme Lagrangian through the minimal coupling prescription,
\begin{equation}\label{cov:U}
    \pmu U\rightarrow D_\mu U = \pmu U - i A_\mu[Q, U],
\end{equation}
where $Q$ is the electromagnetic charge matrix for the $u$ and $d$ quarks,
\begin{equation}
    Q = \begin{pmatrix}
+\frac{2}{3} & 0 \\
0 & -\frac{1}{3}
\end{pmatrix}.
\end{equation}
The elementary charge of the electron has been absorbed into the definition of the gauge field,
\begin{equation}
    A_\mu := \dfrac{e}{\hbar} \mathcal{A}_\mu,\quad [A_\mu] = L^{-1}.
\end{equation}
Replacing the derivatives in \cref{lag:0} by their covariant version, and adding the Maxwell and WZW terms yields the Skyrme-Maxwell model,
\begin{align}
\lden &= \frac{1}{2}\Tr\left(D_\mu U^\dagger D^\mu U\right) - m^2\Tr(\sigma_0 - U) \nonumber \\
      &\quad + \frac{1}{16}\Tr\Big[\big(D_{[\mu} U^\dagger D_{\nu]} U\big)
               \big(D^{[\mu} U^\dagger D^{\nu]} U\big)\Big] \label{lag:D} \\
      &\quad - \frac{a_0^2}{8\pi\alpha}F_{\mu\nu}F^{\mu\nu} + \lden_{\mathrm{\scriptscriptstyle WZW}}. \nonumber
\end{align}
Here $\al \simeq 1/137$ denotes the fine-structure constant and $F_{\mu\nu}$ is the electromagnetic field strength tensor,
\begin{equation}
    F_{\mu\nu} = \partial_\mu A_\nu - \partial_\nu A_\mu.
\end{equation}
The Lagrangian \cref{lag:D} is written in dimensionless units $(x,E)$, which are related to SI units by,
\begin{equation}
    x_{\si{SI}} = \dfrac{2\hbar c}{a_0 f_\pi} x,\quad  E_{\si{SI}} = \dfrac{f_\pi}{4 a_0}  E.
\end{equation}
In these units the pion mass enters through the dimensionless parameter,
\begin{equation}
     m := \dfrac{2 m_\pi c^2}{a_0 f_\pi}.
\end{equation}
The inclusion of the $\lden_{\mathrm{\scriptscriptstyle WZW}}$ term is essential when the electromagnetic field is present. This term represents the Wess–Zumino–Witten contribution \cite{CallanWitten1984}, which accounts for processes induced by QCD anomalies, such as the $\ga\pi^+\pi^-\pi^0$ vertex and the $\pi^0 \rightarrow \ga\ga$ decay, as illustrated in \cref{fig:pion_decay}. The explicit form of the WZW contribution to the Lagrangian is,
\begin{align}
\lden_{\mathrm{\scriptscriptstyle WZW}} &= \dfrac{a_0^2}{8\pi^2} \varepsilon^{\mu\al\be\ga}A_\mu
   \Bigg\{ 2i\,\pal A_\be
   \Tr\bigg[Q^2\{U\dg, \pga U\} \nonumber \\
&\quad +\dfrac{1}{2}Q\left(U\dg Q \pga U -U Q\pga U\dg\right)\bigg] \label{lag:WZW} \\
&\quad -\Tr\bigg[\{Q,U\dg\}\pal U\pbe U\dg\pga U\bigg]\Bigg\}. \nonumber
\end{align}
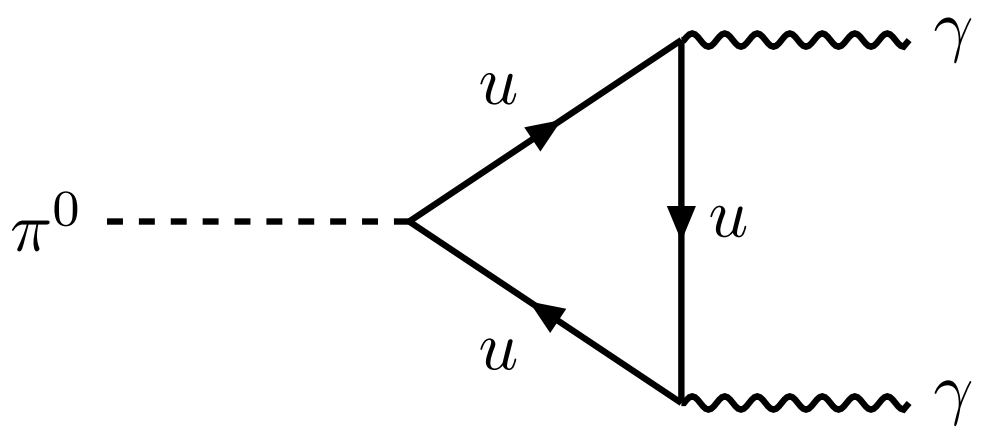
\begin{figure}[h!]
\centering
\begin{tikzpicture}[scale=1.2, transform shape]
\useasboundingbox (-2.5,-1.7) rectangle (3.5,1.7);
\begin{feynman}
    \vertex (pi) at (-2,0) {$\pi^0$};
    \vertex (v1) at (0,0);
    \vertex (v2) at (1.5,1);
    \vertex (v3) at (1.5,-1);
    \vertex (g1) at (3,1) {$\gamma$};
    \vertex (g2) at (3,-1) {$\gamma$};

    \diagram*{
        (pi) -- [scalar, very thick] (v1),

        (v1) -- [fermion, very thick, edge label=$u$] (v2)
              -- [fermion, very thick, edge label=$u$] (v3)
              -- [fermion, very thick, edge label=$u$] (v1),

        (v2) -- [photon, very thick] (g1),
        (v3) -- [photon, very thick] (g2),
    };
\end{feynman}
\end{tikzpicture}
\caption{Neutral pion decay $\pi^0 \to \gamma\gamma$ mediated by the axial anomaly.}
\label{fig:pion_decay}
\end{figure}

As an element of $SU(2)$, the Skyrme field $U$ can be parametrized in terms of four real scalar fields $\phi = (\phi_0, \phi_1, \phi_2, \phi_3)$,
\begin{equation}
    U = \phi_0 \sig_0 + i \phi_k \sig_k,\quad k=1,2,3,
\end{equation}
where we interpret the components $( \phi_1, \phi_2, \phi_3)$ as the pion triplet fields $(\pi^\pm, \pi^0)$, while $\phi_0$ corresponds to the sigma field. The unitarity of $U$ is reflected in the normalization condition,
\begin{equation}
    \phi \cdot \phi = 1,
\end{equation}
which identifies $\phi$ as a point on $S^3$. Inserting the fields into the definition of the covariant derivative \cref{cov:U} gives,
\begin{equation}\label{covder}
    D_\mu \phi_a = \pmu \phi_a + A_\mu \Phi_a,\quad a = 0,1,2,3,
\end{equation}
with the convenient definition,
\begin{equation}
    \Phi_a := \phi_1 \delta_{a2} - \phi_2 \delta_{a1}.
\end{equation}

We are interested in static minimisers of \cref{lag:D}, such that we can assume $\partial_0 \phi = \partial_0 A_\mu = 0$. Under this condition, the static energy density associated with the model \cref{lag:D} is,
\begin{align}
\eden_{\mathrm{tot}} &= D_{ii} + \dfrac{1}{2}\Big[(D_{ii})^2 - D_{ij}D_{ij}\Big] \nonumber \\
      &\quad + A_0^2\Big[(\Phi \cdot D_i\phi)^2 - (\Phi \cdot \Phi)(1 + D_{ii})\Big] \label{eden:static} \\
      &\quad + 2m^2(1-\phi_0) + \dfrac{a_0^2}{8\pi\al}F_{\mu\nu}F^{\mu\nu} + \eden_{\mathrm{\scriptscriptstyle WZW}}, \nonumber
\end{align}
where we have defined the useful tensor,
\begin{equation}
    D_{ij} := D_i\phi \cdot D_j\phi.
\end{equation}
We now seek minimisers of this static energy with a fixed external magnetic field of strength $H$. Without loss of generality we apply this external field in the $z$ direction by fixing the gauge field as,
\begin{equation}\label{ab}
    A_\mu = \dfrac{H}{2} (0, -y, x, 0).
\end{equation}
The gauge invariant electromagnetic fields associated with this four-potential choice is then $\bm{E} = 0$ and $\bm{B} = H \hat{z}$. This leads to the following static energy density that we use for the remainder of this paper,
\begin{equation}\label{eden:H}
\begin{split}
\eden &= D_{ii} + \dfrac{1}{2}\Big[(D_{ii})^2 - D_{ij}D_{ij}\Big]\\
&\quad+ 2m^2(1-\phi_0),
\end{split}
\end{equation}
where the constant Maxwell energy density has been subtracted. We remark that we will set the pion mass $m=0$ in all our numerical calculations. Note that for static fields, only terms proportional to $A_0$, $A_0\,\partial_i A_j$, or $A_i\,\partial_j A_0$ survive in the WZW contribution \cref{lag:WZW} and our choice for the fixed magnetic field ensures that all the remaining terms vanish.

\subsection{Topological current}\label{sec:topo}
Replacing partial with covariant derivatives in the topological current of the ungauged model yields,
\begin{equation}\label{B_unga}
    B_g^\mu = - \dfrac{1}{24 \pi ^2 } \varepsilon^{\mu\al\be\ga} \Tr (U\dg D_\al U D_\be U\dg D_\ga U),
\end{equation}
or in terms of the ungauged baryonic current $B^\mu$,
\begin{equation}
    B_g^\mu = B^\mu + \frac{i}{8\pi^2}\varepsilon^{\mu\alpha\beta\gamma} A_\ga\Tr (\pal U\dg \pbe U \Tilde{Q}),
\end{equation}
with $\Tilde{Q} = (U\dg Q U - Q)$. The quantity $B_g^\mu$ is gauge invariant but not conserved,

\begin{equation}
    \partial_\mu B_g^\mu = \frac{1}{2\pi^2}\,\tilde F^{\mu\nu}\,\partial_\mu\phi_0\,\partial_\nu\phi_3.
\end{equation}
We seek an extra term that compensates this four-divergence. A truly gauge invariant and conserved current is,
\begin{equation}
     \mathcal{B}^\mu = B^\mu_g + \frac{1}{2\pi^2}\,
    \tilde F^{\mu\nu}\,\phi_3\,\partial_\nu\phi_0.
\end{equation}
This choice differs from that of Callan and Witten \cite{CallanWitten1984} by a term $\propto \pnu(\tilde F^{\mu\nu} \phi_0\phi_3)$ that is identically conserved. The zeroth component of $\mathcal{B}^\mu$ gives,
\begin{equation}
     \mathcal{B}^0 = B^0 + \dfrac{1}{2\pi^2} \varepsilon^{ijk} (\partial_i\phi_3 A_j + \phi_3 \partial_i A_j)\partial_k\phi_0,
\end{equation}
where the terms added to $B^0$ can be written as a total divergence,
\begin{equation}
    \mathcal{B}^0=B^0+\partial_i\Omega_i,\quad \Omega_i=\frac{1}{2\pi^2}\,\varepsilon^{ijk}A_j\,\phi_3\,\partial_k\phi_0.
\end{equation}
Integrating this topological density over space gives
\begin{equation}\label{topoint}
    \mathcal{B} = B + \oint_{\partial V}\Omega_i\,dS_i.
\end{equation}
At large distances $r\to\infty$, the solution for the $\phi$ fields from the linearized equations of motion gives the asymptotic behaviour $\Omega\sim r^{-3}$. The surface term in \cref{topoint} vanishes and the gauged topological number coincides with that of the ungauged model,
\begin{equation}
    \mathcal{B} = B.
\end{equation}

The baryon number normally sets a lower bound for the energy \cite{Piette2000}. The bound for the $A_0=0$ and $m=0$ case of the total energy density \cref{eden:static} is,
\begin{equation}\label{piettebound}
    \eden_{\mathrm{tot}} \geq  12 \pi^2 \left( 1+\dfrac{9\pi \al}{a_0^2} \right)^{-1/2}  B^0.
\end{equation}
However, in the constant magnetic field scenario, spatial integration of \cref{piettebound} gives a trivial inequality, since the left part contains the following free field term that integrates to infinity,
\begin{equation}\label{eden:em}
    \eden_{em} =  \dfrac{a_0^2}{4 \pi \al} H^2.
\end{equation}
Hence, the standard approach doesn't provide a meaningful bound. Nevertheless, we find that energy curves present a global minimum, owing to a balance in the other energy terms.

\subsection{Derrick scaling}

We seek local minimizers $\phi_m$ of \cref{eden:H} which are critical w.r.t. all deformations of $\phi_m$. Hence, an isotropic spatial rescaling $x_i \to \lambda x_i$ corresponds to a critical point when $\lambda = 1$. Historically, the Skyrme term was introduced to ensure such a critical point exists \cite{Derrick1964}, stabilising the soliton against collapse by shrinking.

In our system, as the external magnetic field is fixed, the covariant derivative exhibits a mixed scaling behaviour,
\begin{equation}
    D_i \phi_a 
    = \underbrace{\partial_i \phi_a}_{\sim \lambda^{-1}} 
    + \underbrace{\frac{H}{2} (\hat{z} \times \Vec{x})_i \Phi_a}_{\sim \lambda}.
\end{equation}
We can therefore write the energy \cref{eden:H} associated with the spatial rescaling $\lambda$ of a general configuration $\phi$ as,
\begin{equation}
    E(\lambda) = \lambda^{-1} E_{4} + \lambda E_{2} + \lambda^{3} E_0 + \lambda^{5} E_{-2},
\end{equation}
where $E_i$ are energy terms with the subscript indicating the order of spatial dependence of the energy density. Explicitly, these terms are,
\begin{align}
E_{4\hphantom{-}} &= \dfrac{1}{2}\int d^3x\, \big[ d_{ii}^2 - d_{ij}d_{ij} \big], \\[8pt]
E_{2\hphantom{-}} &= \int d^3x\, \big[ d_{ii} + 2\left( d_{ii}A_j - d_{ij}A_i \right) k_j\big], \\[8pt]
E_{0\hphantom{-}} &= \int d^3x\, \big[ 2m^2(1-\phi_0) +\ 2A_ik_i \nonumber \\
                      &\quad +\ (A_ik_i)^2 - A_iA_i\,k_jk_j \\
                      &\quad +\ \left( d_{ii}A_jA_j - d_{ij}A_iA_j \right)(\phi_1^2 + \phi_2^2)\big], \nonumber\\[8pt]
E_{-2}            &= \int d^3x \big[A_iA_i\,(\phi_1^2 + \phi_2^2)\big],
\end{align}
where $
    d_{ij} := \partial_i \phi \cdot \partial_j\phi$ and
    $k_i := \Phi \cdot \partial_i \phi$,
leading to the virial equation for a critical point $\phi_m$,
\begin{equation}\label{derrick_energy}
\left.\frac{d E[\phi_m](\lambda)}{d\lambda}\right|_{\lambda = 1}\!\! = -E_{4} + E_{2} + 3 E_0 + 5 E_{-2} = 0.
\end{equation}
Satisfying this relation between the energy terms is a useful test to ensure the results of our numerical simulations are local minima $\phi_m$ up to numerical error.

\section{Numerical Method}\label{sec:method}

We prepared a regular discretization of space using a 3D-lattice of $N^3$ points with spacing $h$, where we used $N = 280$ and $h = 0.1$. This allowed the fields to relax to the vacuum value $\phi_{vac} = (1,0,0,0)$ within numerical error before reaching the boundary, where we impose Dirichlet boundary conditions ($\phi = \phi_{vac}$). Spatial derivatives were computed using a fourth-order central finite difference scheme. Given the highly parallelizable nature of the problem, the algorithm was implemented on a GPU using CUDA.

To approximate local minimisers of the system we adopt an arrested Newton flow (ANF) method \cite{speight2020}, a second order method that accelerates the flow in the direction of steepest decent,
\begin{equation}
    \ddot{\phi} + \dfrac{\delta\eden}{\delta\phi} = 0.
\end{equation}
In practice, as we represent $\phi$ as a $4$-vector, the second-order Newton equation above is evolved as a coupled system of projected first-order ODEs,
\begin{equation}\label{ODEsystem}
  \dot{\psi} = -P_\phi\!\left(\dfrac{\delta\eden}{\delta\phi}\right), \quad
  \dot{\phi} = P_\phi(\psi),
\end{equation}
where $\psi$ is an auxiliary velocity field and the projection $P_\phi$ ensures that the gradient and $\psi$ are in the tangent space of $S^3$ at $\phi$,

\begin{equation}
P_\phi (X) := X - (X\cdot\phi) \phi.
\end{equation}

The ODE system \cref{ODEsystem} is integrated iteratively from an initial guess within the desired topological sector $B$. This initial configuration is either obtained from the rational map or from a previously converged solution corresponding to a different value of the parameter $H$. The iteration was performed using an explicit 4th-order Runge-Kutta method, normalising $\phi$ after each step. At each iteration we monitor whether the velocity remains aligned with the downhill direction of the gradient. Whenever,
\begin{equation}
    \dot{\psi} \cdot \psi < 0,
\end{equation}
the velocity is directed away from the gradient. In this scenario, we arrest the momentum of the system by setting $\dot{\psi} = 0$. A numerical solution is considered converged once the magnitude of the gradient at every grid point is below a tolerance of $10^{-5}$. We gauge the accuracy of a given numerical solution as how close the virial relation in \cref{derrick_energy} is to vanishing,
\begin{equation}
    \Delta E = |E_{4} - E_{2} - 3 E_0 - 5 E_{-2}|.
\end{equation}
This quantity is also useful for tuning the grid parameters: if the lattice spacing is too wide to resolve the field details, $\Delta E$ will be large. If the box size is too small, the boundaries will artificially constrain the solution, increasing the value of $\Delta E$. The error in our solutions is primarily due to finite-size effects induced by the the walls and only affects some of the most spatially extended configurations. Even in these cases, it remains below approximately $0.3\%$ of the total energy per baryon number.

\section{Numerical results}\label{sec:results}

\begin{figure*}[p]
  \centering
  \includegraphics[width=\linewidth]{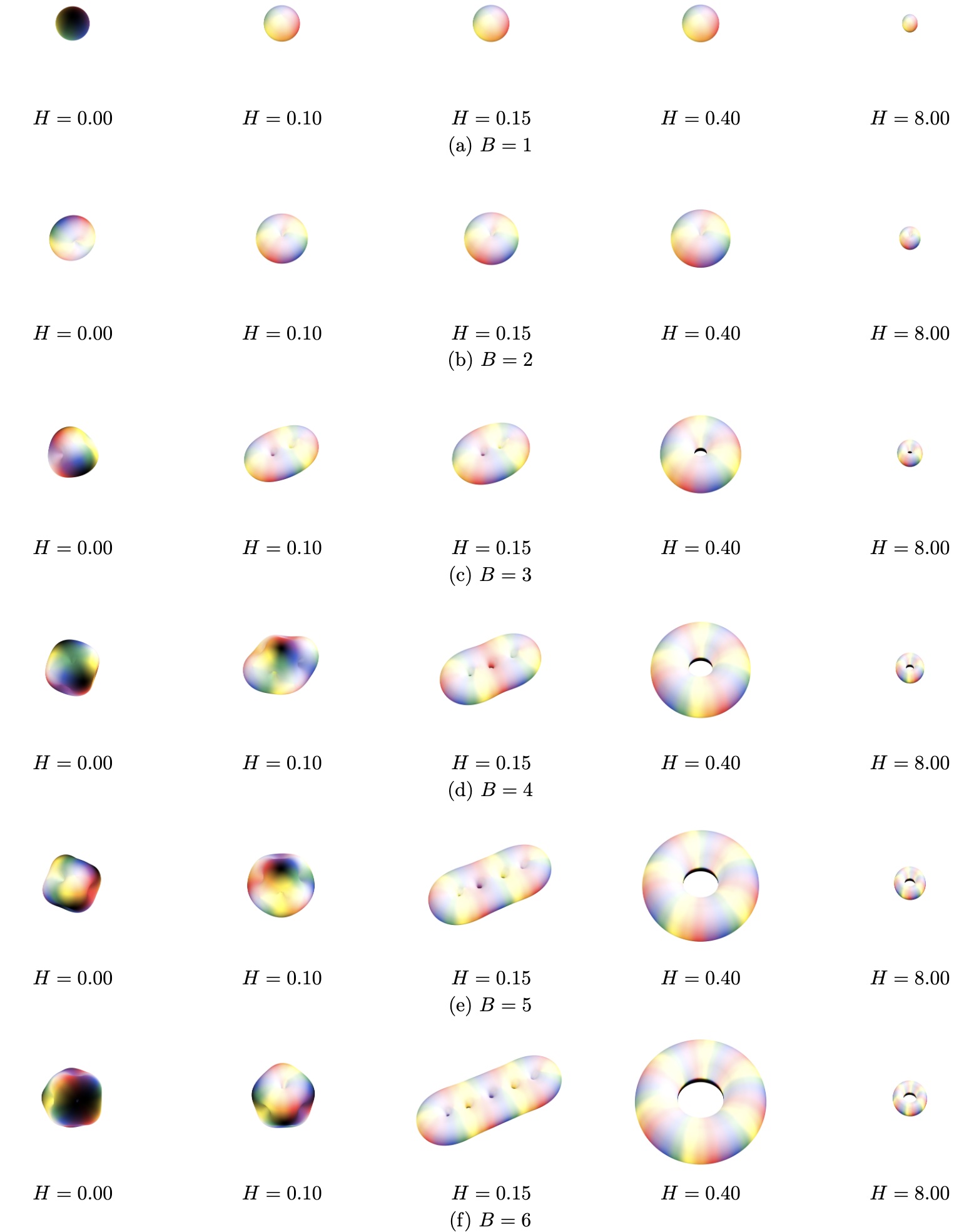}
    \caption{Minimal-energy baryon density isosurfaces versus the external magnetic field $H$. All figures are shown at the same scale, such that relative sizes are informative. }
    \label{fig:all_in_one}
\end{figure*}

In this section we present the numerical results for massless skyrmions coupled to a constant external magnetic induction $H \hat{z}$. We find qualitative features common to all baryon numbers $B=1$–$6$. In general, the energy of the static solutions is lower than in the ungauged model, the whole energy landscape is discussed in \cref{sec:energy}.

A wide view of the coupled soliton configurations is shown in \cref{fig:all_in_one}, while their visual representation is explained in appendix~\ref{sec:visual_rep}. This visualisation tells us that both the baryon density distribution and size of the soliton are affected by the magnetic field.  When $H$ is below a certain value, we observe that the gauged skyrmion is larger than the ungauged one. In contrast, a stronger magnetic field reduces the size of the configuration. In the $B=1$ and $B=2$ sectors, the baryon density is a smooth deformation of the ungauged rational map geometry for all values of $H$. This allows us to analise the shape and extension of these solutions in greater detail in \cref{b1b2}, finding that the higher the external field is, the more prolate the solution becomes.

For $B>2$ other geometrical configurations different from the rational-map shape may minimize the energy. We breakdown the possible topologies and the transition between them in \cref{sec:b>2}. Indeed  we find that, for small field, the static solution is a slight deformation of the $H=0$ configuration, but when the induction is strong, the baryon density becomes a $g$-toroid with $g=B-1$. For even stronger values of $H$, the geometry becomes a simple toroid. In the transition region between the deformed rational map shape and the $g$-toroid, other geometries with similar energy appear and become the energy minimisers for small intervals of $H$.

In the sectors $B=3, 5, 6$ we find folded structures in the rational-map like distributions that provide negative baryon density regions, whereas the baryon density of all the other geometries is non-negative everywhere. This is discussed in \cref{sec:neg_density}.

\subsection{Energy profile of \texorpdfstring{$B\leq 6$}{B<=6} gauged skyrmions}\label{sec:energy}

\begin{figure}[H]
  \centering
  \includegraphics[width=\linewidth]{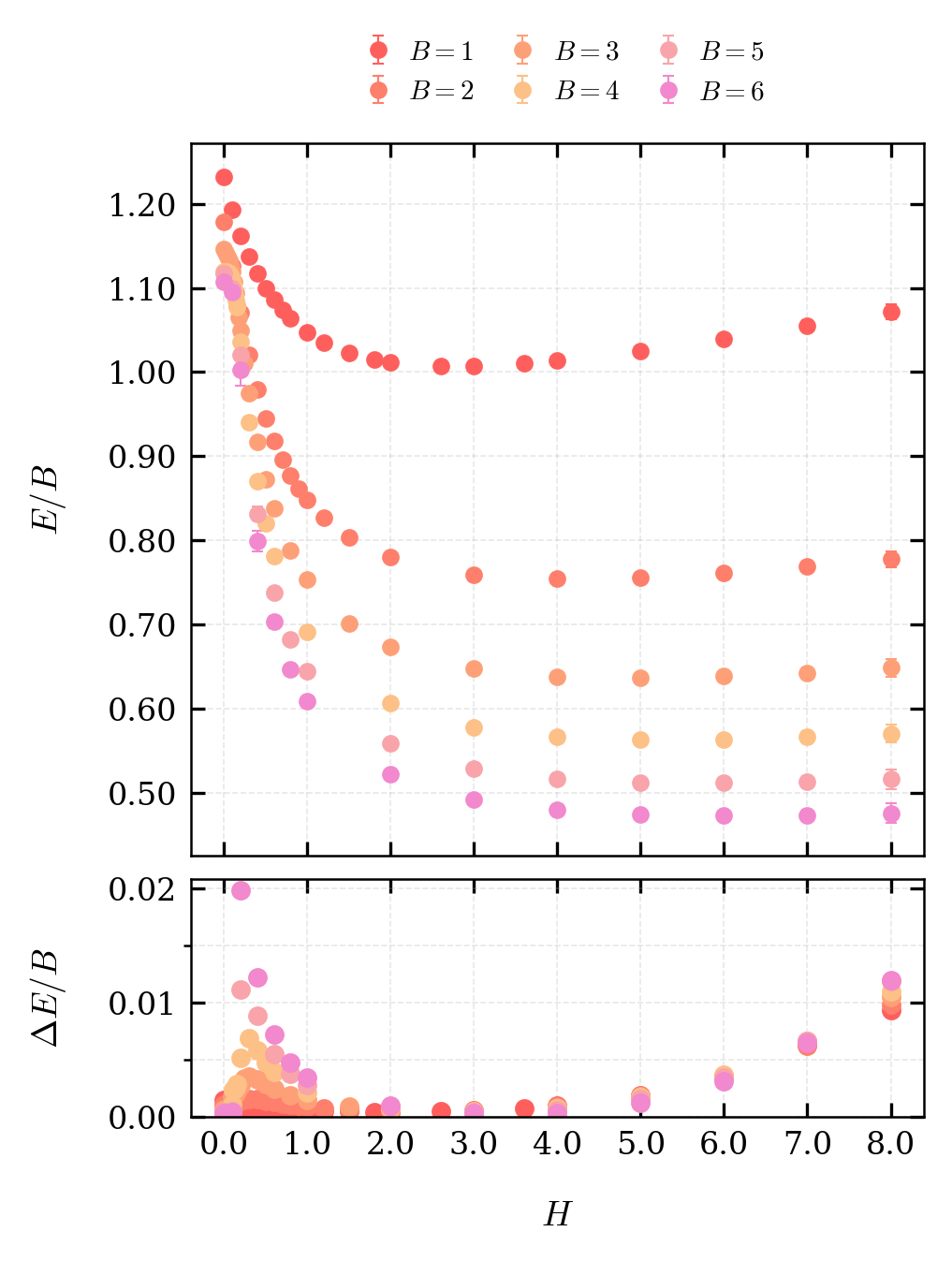}
  \caption{Energy per baryon number versus external magnetic field for $B \leq 6$.}
  \label{fig:energy}
\end{figure}

We want to find field configurations that minimize the massless Skyrme-Maxwell energy,
\begin{equation}\label{eint}
E(H) = \int d^3x \bigg[ D_{ii} + \dfrac{1}{2} \bigg((D_{ii})^2 -D_{ij}D_{ij} \bigg) \bigg].
\end{equation}
For a given value of $H$, we seek solutions using the method described in \cref{sec:method}. We cover the range $H\in [0.0, 8.0]$. The energy profile for baryon numbers one through six is shown in \cref{fig:energy}.
We observe that for all magnetic field values, increasing the baryon number leads to more stable configurations in terms of energy per baryon unit. This energy difference, relative to $B=1$, is interpreted as the binding energy of the configuration. The energy curves have a global minimum $H_{min}$ that changes with $B$. They decrease sharply for $H<H_{min}$, while it slowly increases for $H>H_{min}$. The existence of $H_{min}$ is not related to any known topological bound. Instead, it is explained from the balance between energy terms that scale with different powers of $H$ in \cref{eden:H}. The energy increase at extremely high magnetic fields is associated with the compression of the skyrmion. However, this compression alone cannot fully account for the energy profile. As we will show in the next section, the field strength $H$ at which the skyrmion reaches its lowest density does not coincide with the value that minimises its energy.

\subsection{Deformation of $B=1$ and $B=2$}\label{b1b2}
\begin{figure}[htbp]
    \centering
    \begin{tabular}{@{}ccc@{}}
        \includegraphics[width=0.3\linewidth]{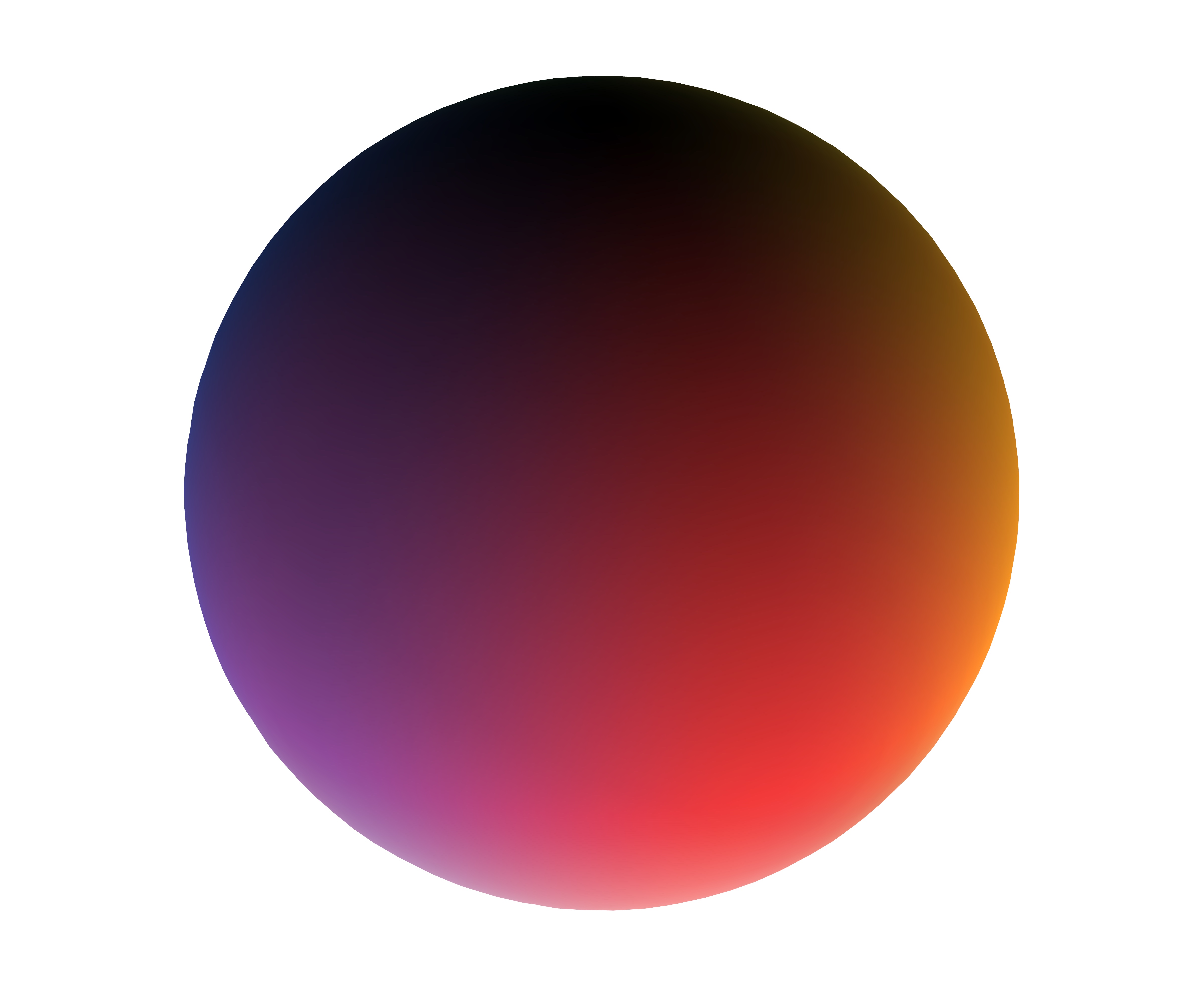} &
        \includegraphics[width=0.3\linewidth]{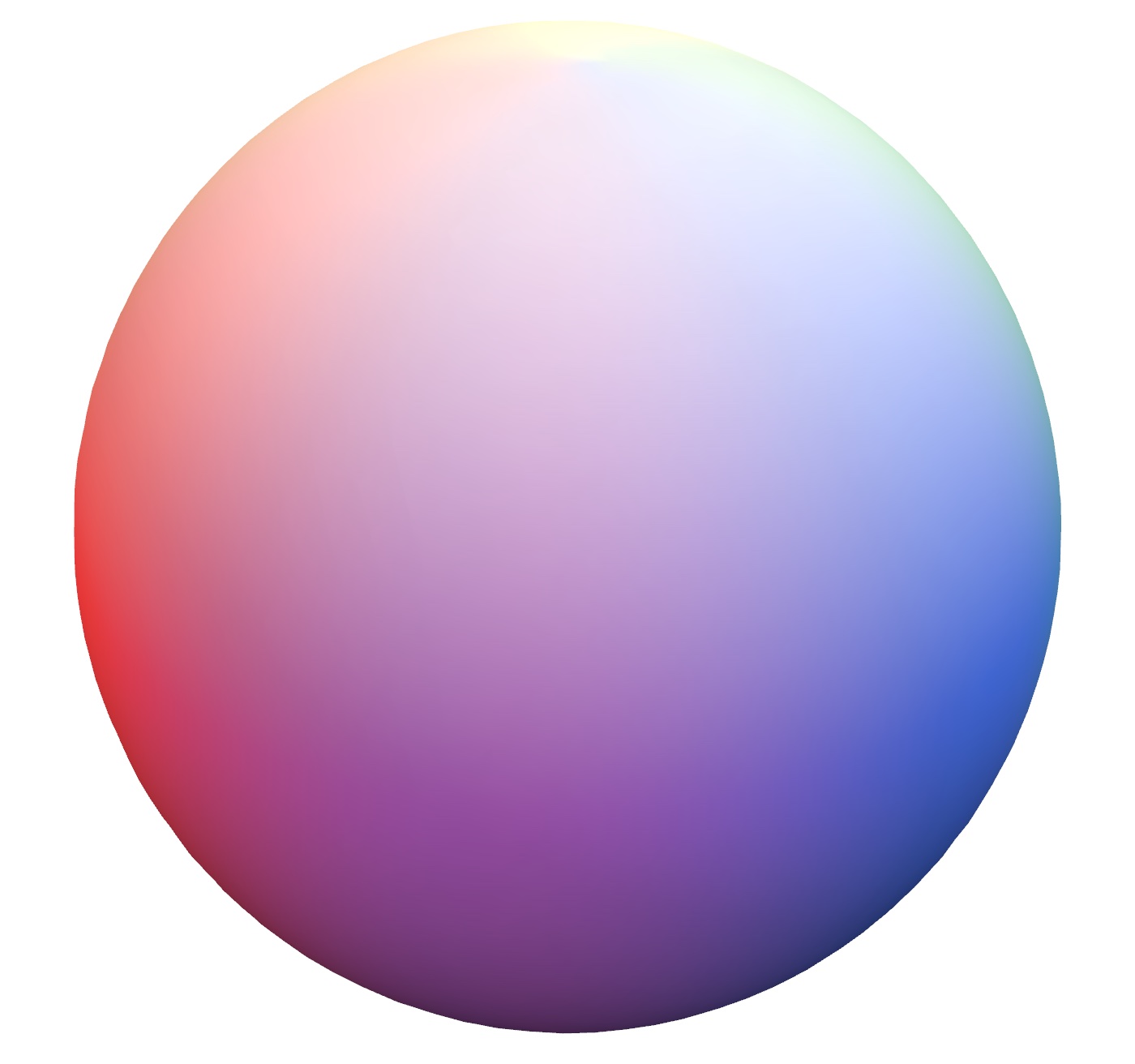} &
        \includegraphics[width=0.3\linewidth]{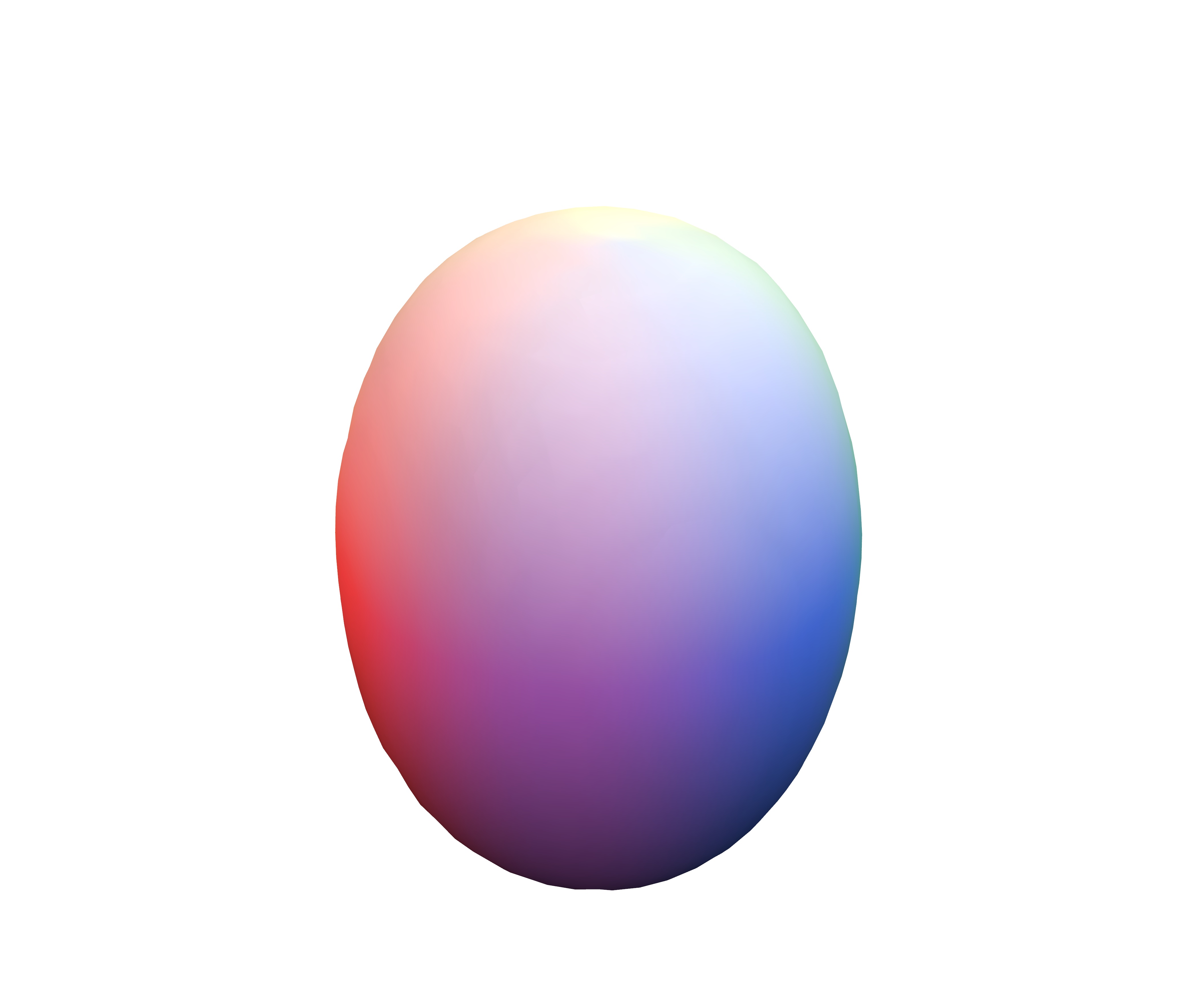} \\
        \small $H=0.0$ & \small $H=0.30$ & \small $H=5.0$
    \end{tabular}
    \caption{Baryon density isosurfaces for $B=1$.}
    \label{fig:B1_configurations}
\end{figure}

\begin{figure}[htbp]
    \centering
    \begin{tabular}{@{}ccc@{}}
        \includegraphics[width=0.3\linewidth]{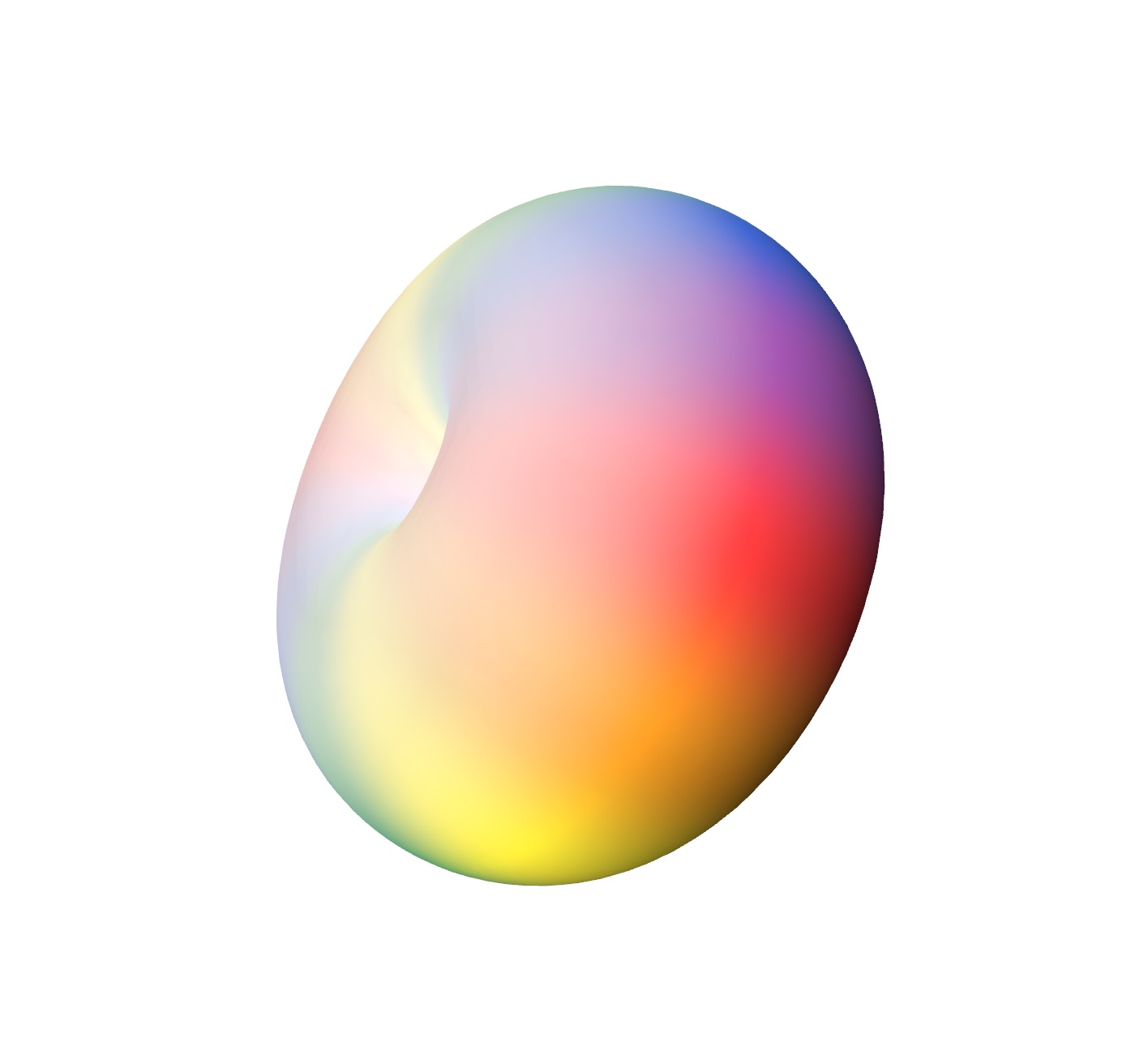} &
        \includegraphics[width=0.3\linewidth]{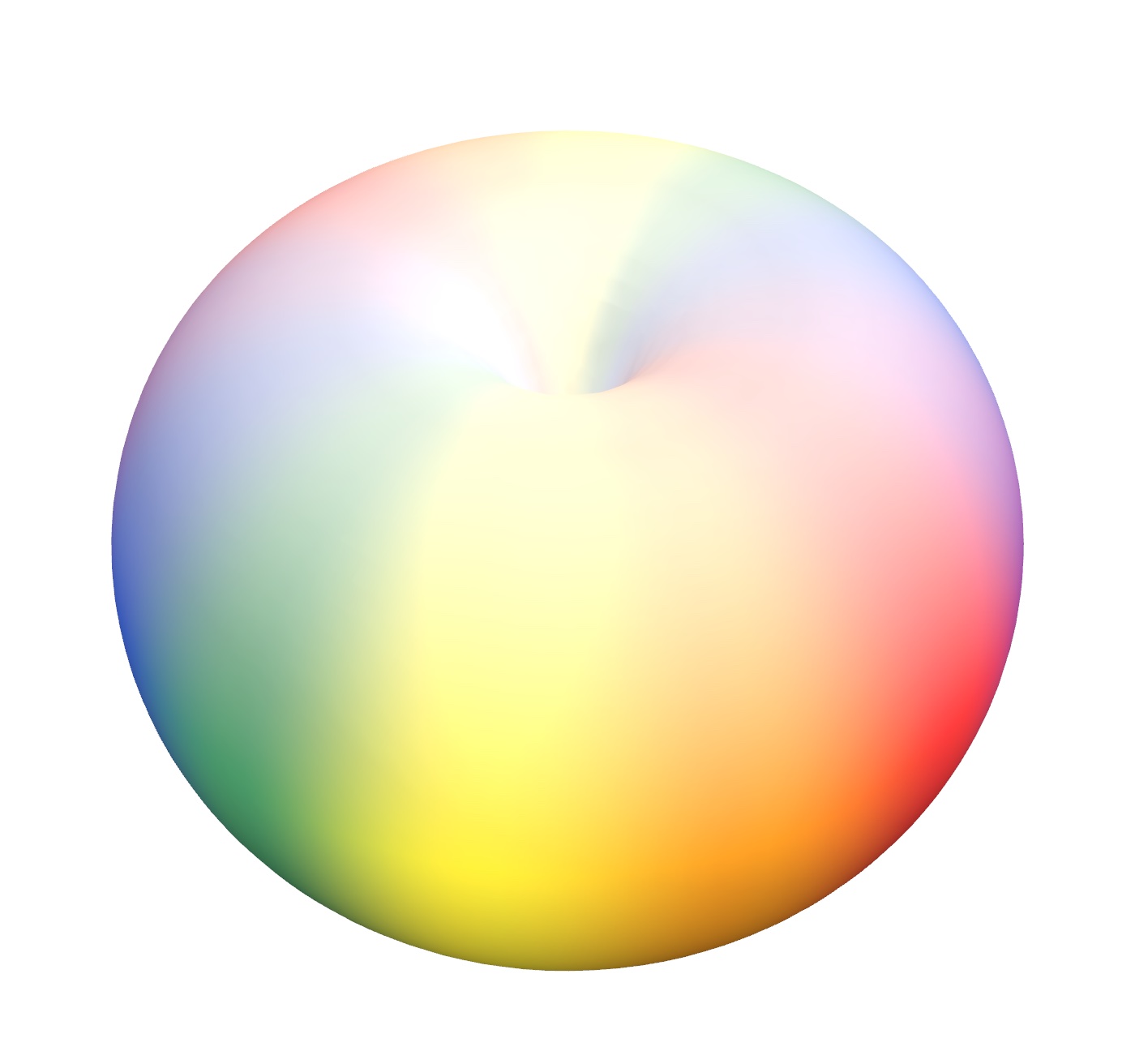} &
        \includegraphics[width=0.3\linewidth]{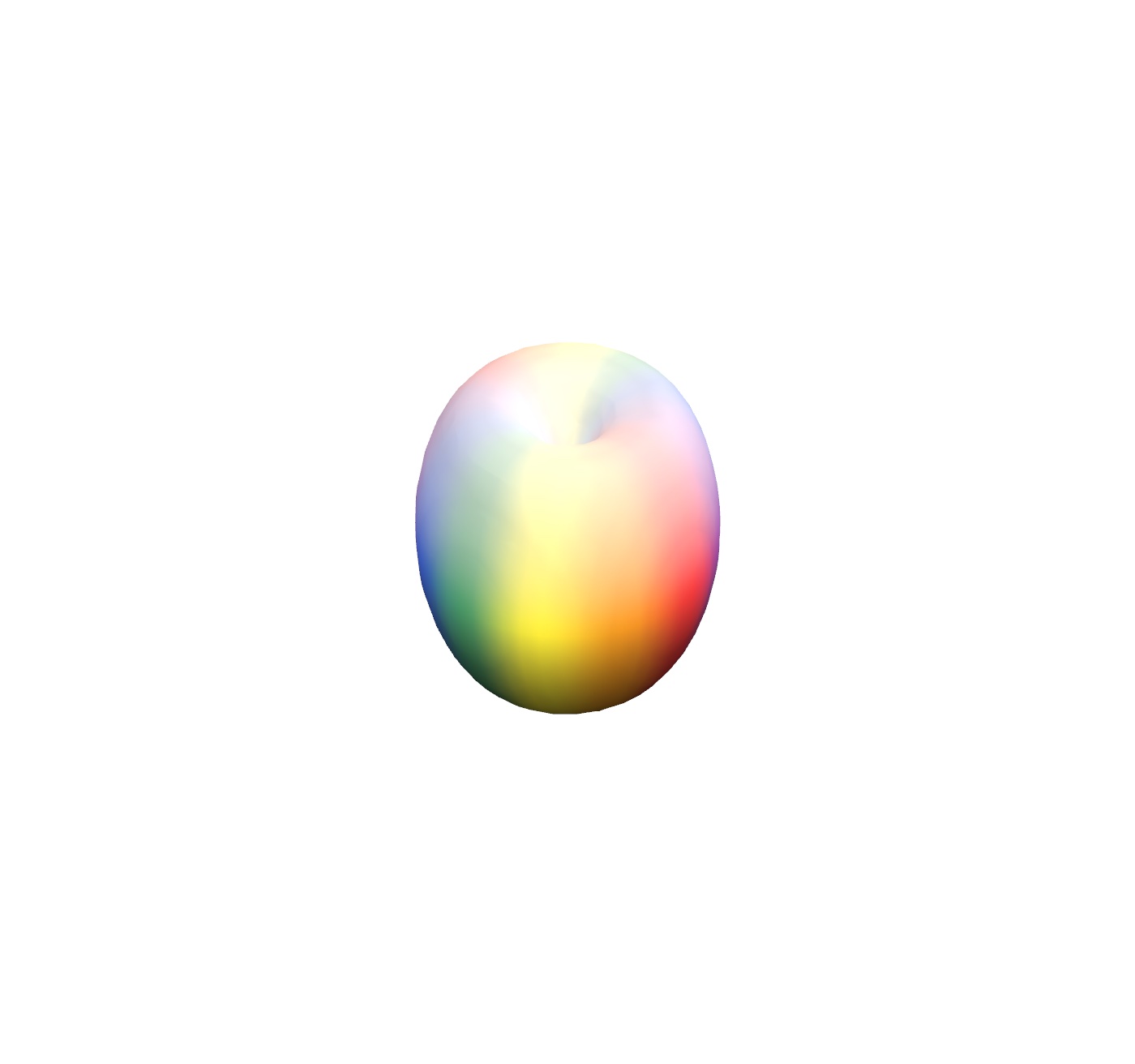} \\
        \small $H=0.0$ & \small $H=0.40$ & \small $H=8.0$
    \end{tabular}
    \caption{Baryon density isosurfaces for $B=2$.}
    \label{fig:B2_configurations}
\end{figure}

In the $B=1,2$ sectors, the geometry of the gauged soliton is a smooth deformation of the ungauged stable configuration. These deformations correspond to a stretching along the direction of the external field, of the spherical and toroidal shapes given by the rational map. The baryon density isosurfaces, which enclose the $90\%$ of the total baryon charge, are displayed in \cref{fig:all_in_one,fig:B1_configurations,fig:B2_configurations}. Note that the axis of rotational symmetry aligns with the magnetic field direction $\hat{z}$.

\begin{figure}[h!]
  \centering
  \includegraphics[width=\linewidth]{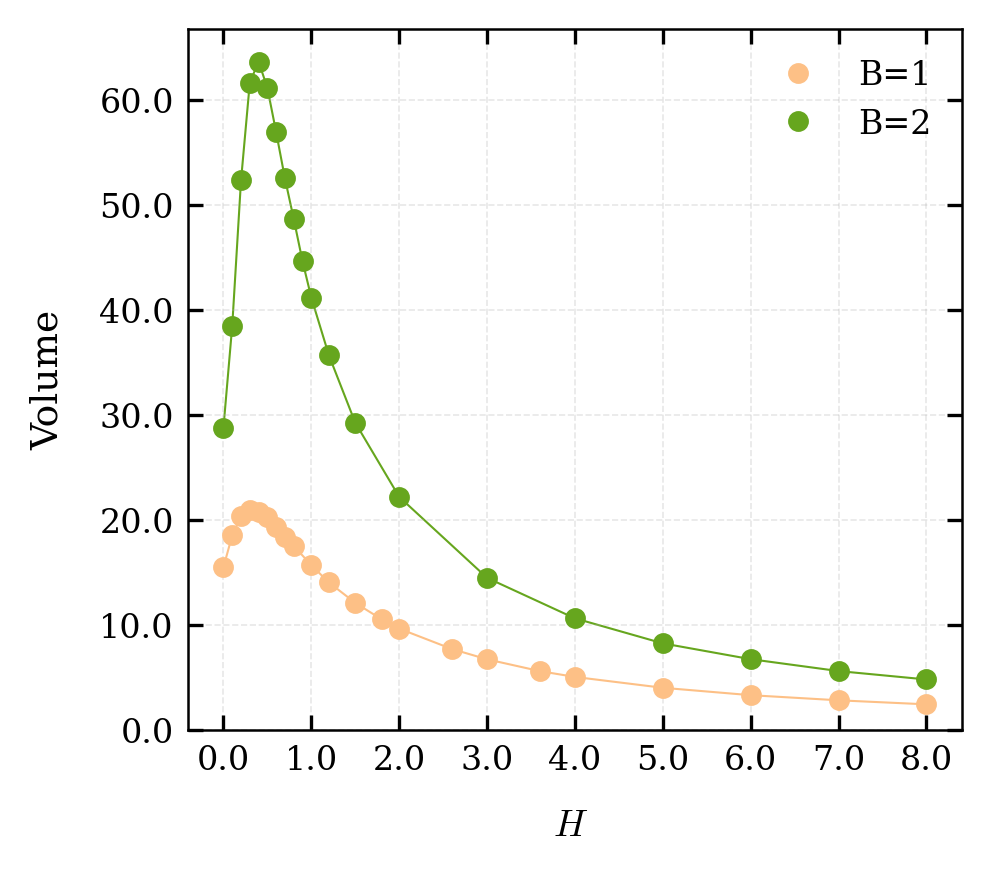}
  \caption{Volume enclosing the 90\% of the baryon charge.}
  \label{fig:volume}
\end{figure}

We evaluate the volume enclosed by these isosurfaces and examine its dependence on the magnetic field. The result of this measurement is shown in \cref{fig:volume}. The volume initially increases with $H$, reaching a maximum at field value $H_c \simeq 0.4$. For $H > H_c$, the volume decreases monotonically as the magnetic field is further increased. The value $H_c$ is not related with the magnetic field that gives the minimum of the energy curves in \cref{fig:energy}.

\begin{figure}[H]
  \raggedleft
  \begin{subfigure}[b]{\linewidth}
    \centering
    \includegraphics[width=\linewidth]{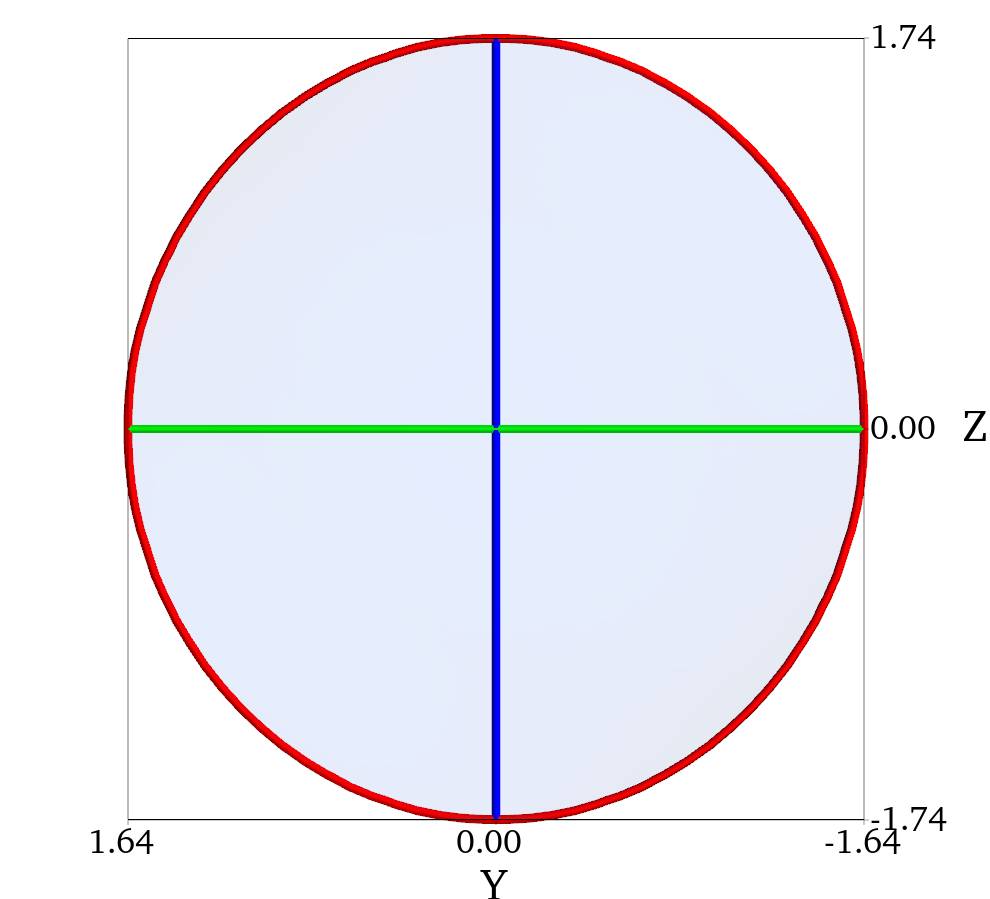}
    \caption{$B=1$}
    \label{fig:B1_cut}
  \end{subfigure}
  \par\medskip
  \begin{subfigure}[b]{\linewidth}
    \centering
    \includegraphics[width=1\linewidth]{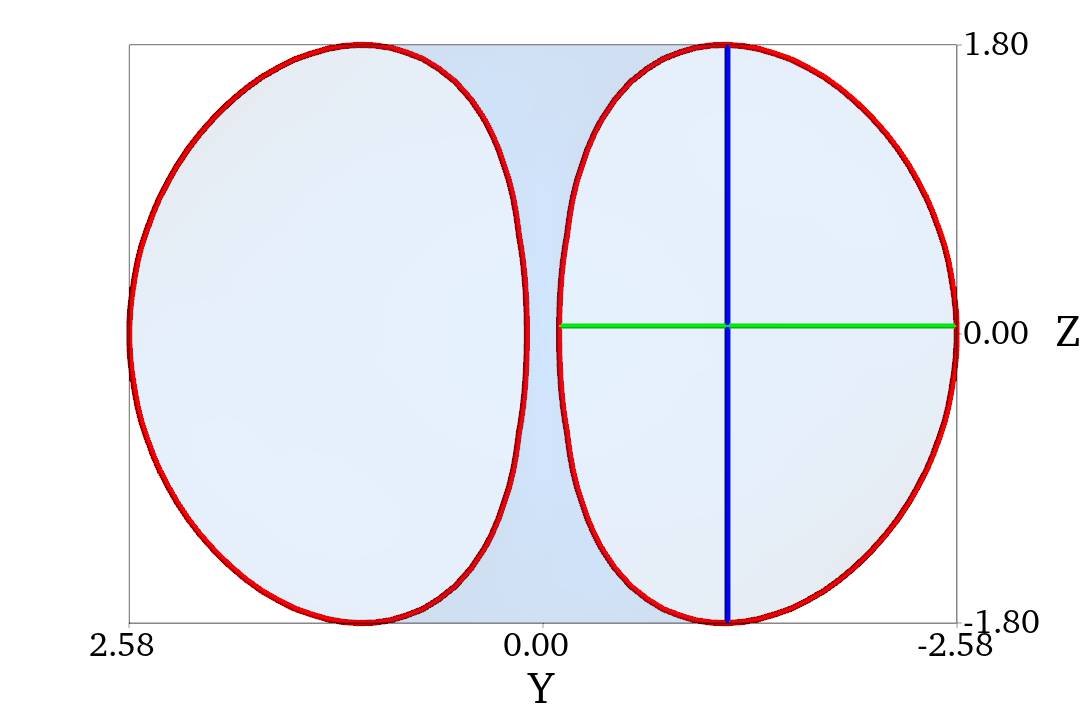}
    \caption{$B=2$}
    \label{fig:B2_cut}
  \end{subfigure}
  \caption{Cut of the baryon density isosurfaces with the $yz$-plane at $H=0.60$. The green (blue) segment marks the minimum (maximum) width $\Delta y$ ($\Delta z$).}
  \label{fig:cuts}
\end{figure}

\begin{figure}[H]
  \centering
  \includegraphics[width=\linewidth]{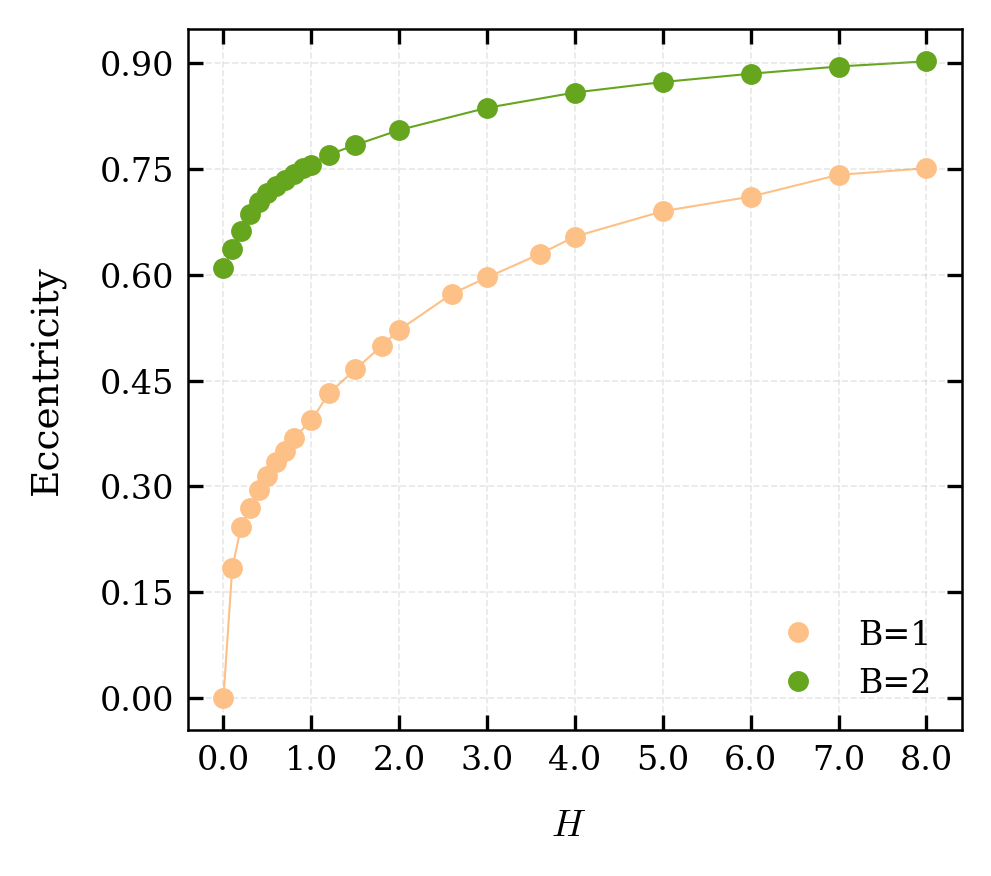}
  \caption{Eccentricity of the baryon density isosurfaces.}
  \label{fig:eccentricity}
\end{figure}

Not only does its size change, but the soliton is also stretched along the direction of the field. The gauged $B=1$ skyrmion adopts a revolution ellipsoid-like shape \cref{fig:B1_configurations}, while the $B=2$ configurations have an elongated toroidal structure \cref{fig:B2_configurations}. Taking advantage of their axial symmetry, we can examine the intersection of the isosurfaces with the yz-plane. This cut gives a simple, closed, convex curve for $B=1$ and two such curves for $B=2$, as shown in \cref{fig:cuts}. We can measure their eccentricity with the quantity,
\begin{equation}
    e = \sqrt{1-(\Delta y/\Delta z)^2},
\end{equation}
where $\Delta y$ ($\Delta z$) is the minimum (maximum) distance between two points of the curve, along a direction perpendicular to the curve. We observe that $\Delta z$ always corresponds to the width along the $\hat{z}$ direction, whereas $\Delta y$ corresponds to the width in the perpendicular direction. This confirms that the magnetic field stretches the soliton along its direction. Moreover, the stronger the external field is, the more prolate the isosurface becomes. This increase of the eccentricity with the magnetic field is shown in \cref{fig:eccentricity}.

It has been found that similarly deformed solutions for $B=1,2$ can be found by introducing an anisotropic potential, breaking the isospin symmetry of the ungauged model \cite{winyard2016skyrme}. The external magnetic field plays the role of breaking this symmetry in our Skyrme-Maxwell model.

\subsection{Configurations for $B>2$ }\label{sec:b>2}

\begin{figure}[h!]
  \centering
  \includegraphics[width=\linewidth]{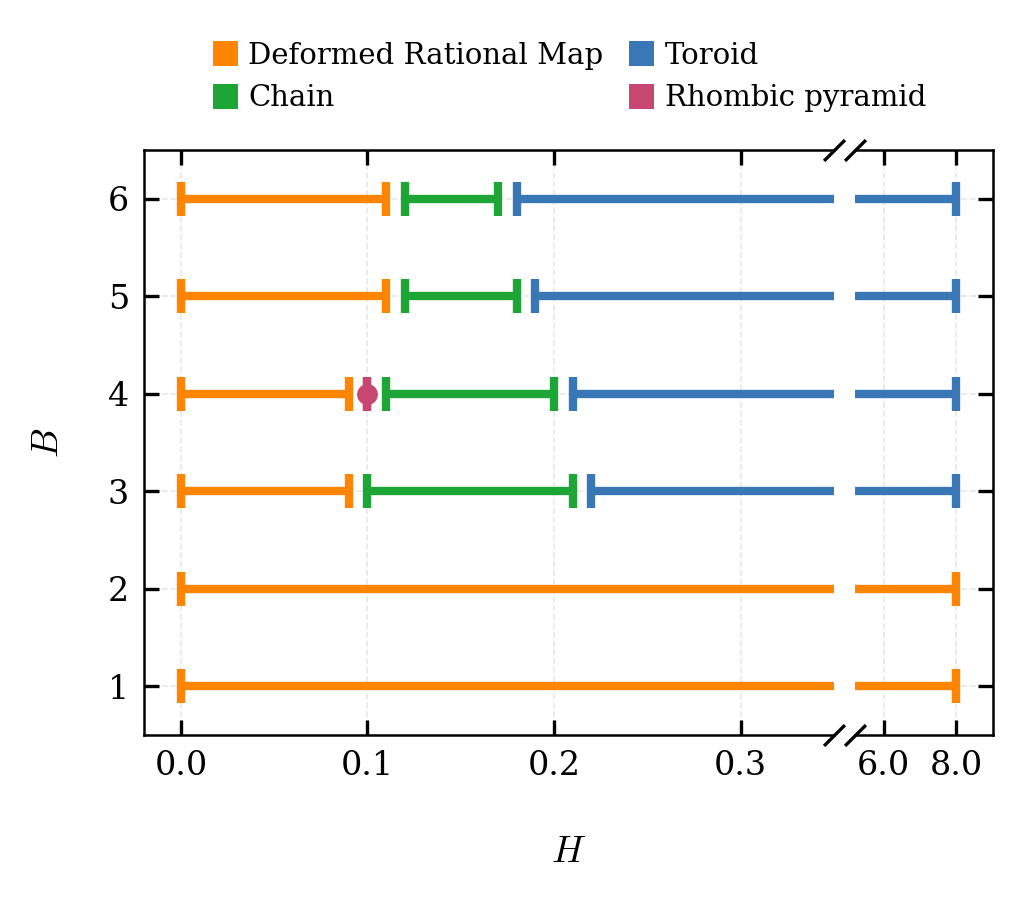}
  \caption{Ranges of $H$ over which each baryon density geometry minimizes the energy.}
  \label{fig:minimizer_intervals}
\end{figure}

As discussed above, the $B=1$ and $B=2$ static solutions of the Skyrme-Maxwell model correspond to smooth deformations of the ungauged skyrmions. However, for B>2 gauged skyrmions, a variety of geometrical shapes compete for the minimum energy state. Depending on the strength of $H$, we identify two distinct regimes sharing the same qualitative features across all baryon numbers:

\begin{itemize}
    \item \textit{Weak field.} The solutions retain a slightly deformed rational-map shape. As the external field increases, new configurations emerge as competing local minima, driving transitions between different shapes.
    
    \item \textit{Strong field.} Energy minimizers become more predictable. Within a narrow range of $H$, they form a $g$-toroidal structure with $g= B-1$, that is aligned in a straight chain. For higher values of the magnetic field, the isosurfaces become a simple $1$-toroid.
\end{itemize}
The specific ranges of $H$ associated with these two regimes depend on the baryon number. In \cref{fig:minimizer_intervals} we illustrate the corresponding intervals and the energy-minimising configurations within each of them. In the following, we analyse these two regimes in detail.

\subsubsection{Weak magnetic field transitions}
\begin{figure}[h!]
     \centering
     \begin{subfigure}[b]{0.49\linewidth}
         \centering
         \includegraphics[width=\linewidth]{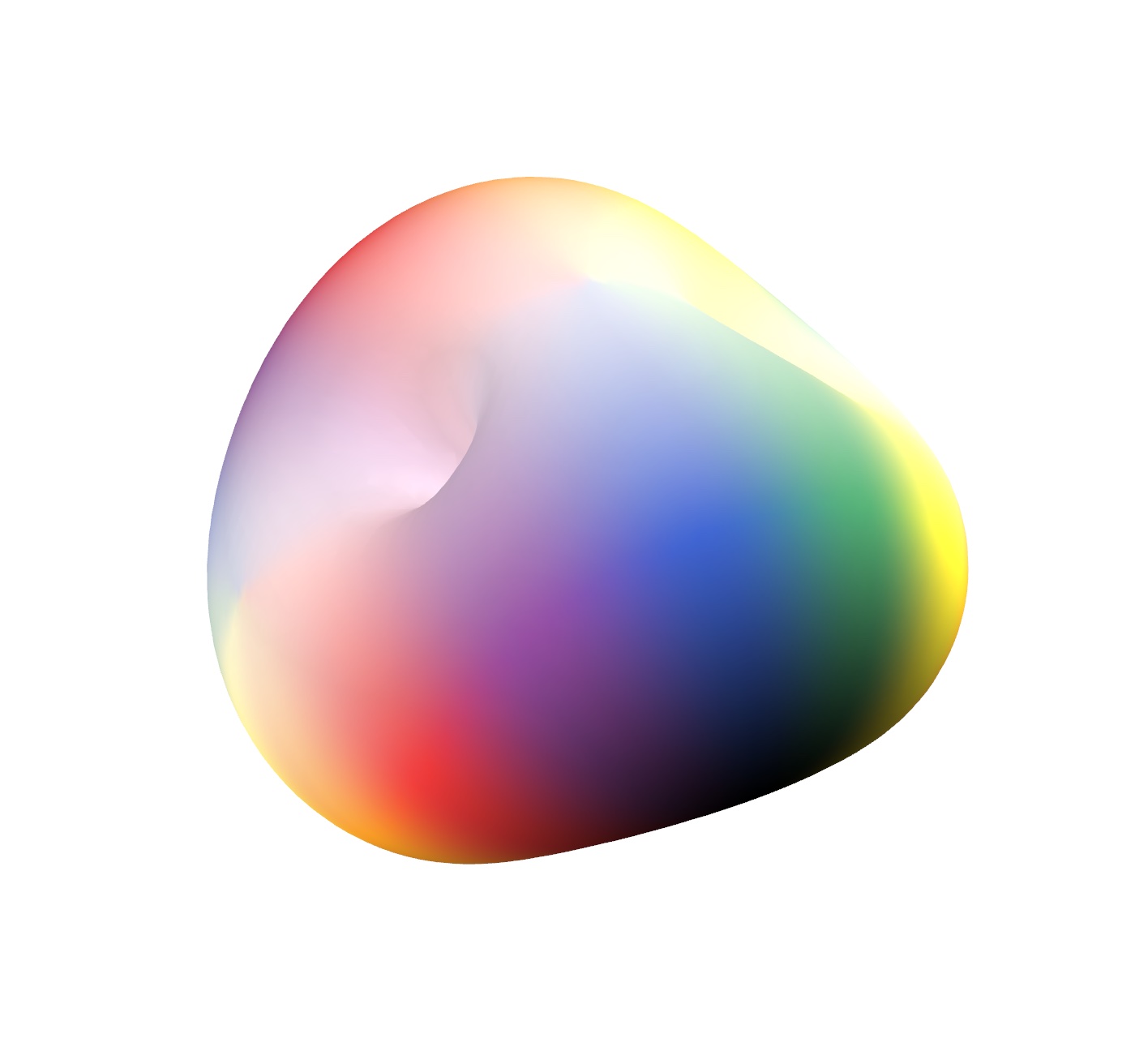}
         \caption{Tetrahedron}
     \end{subfigure}
     \hfill
     \begin{subfigure}[b]{0.49\linewidth}
         \centering
         \includegraphics[width=\linewidth]{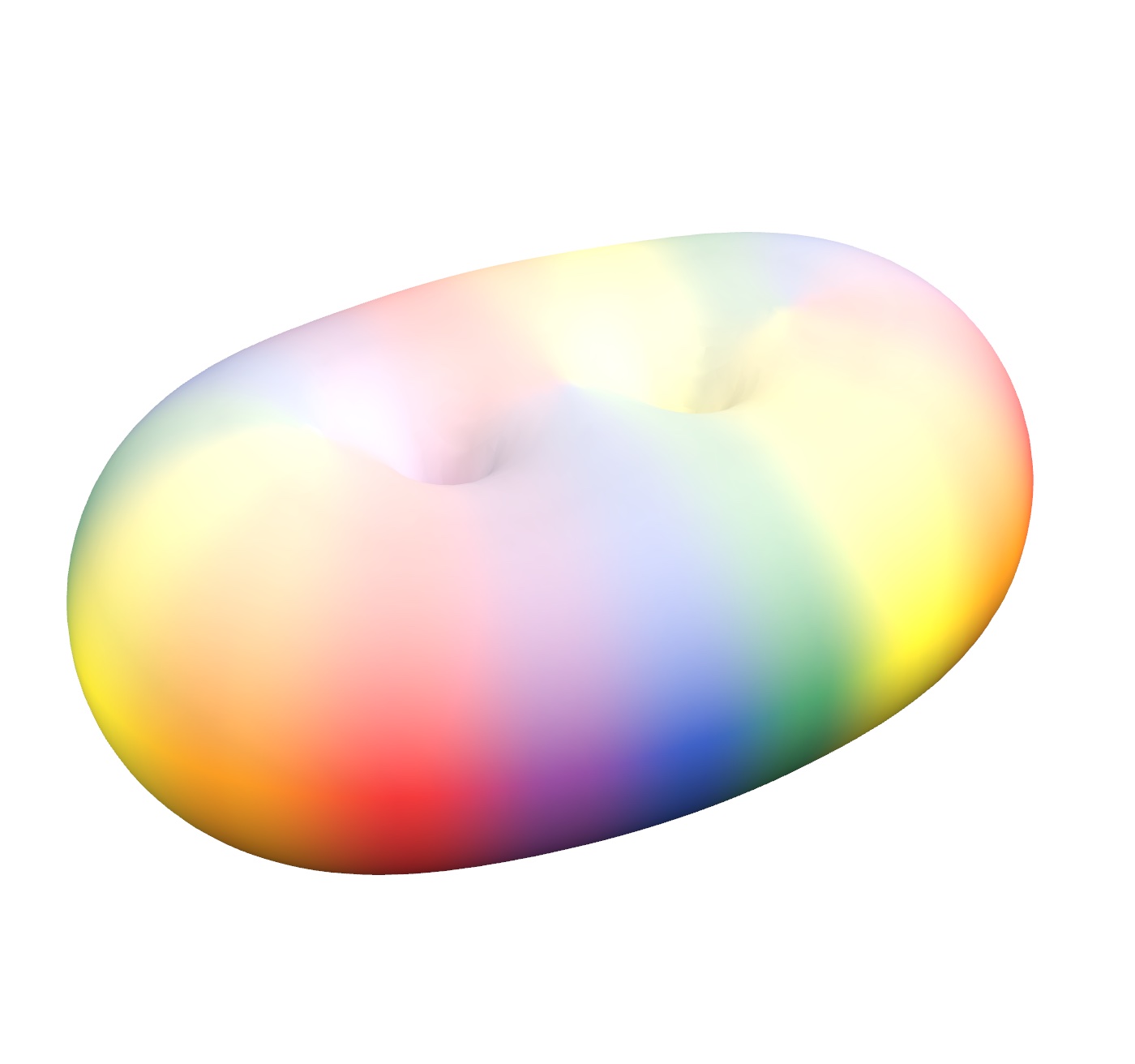}
         \caption{Chain}
     \end{subfigure}
     
    \caption{Competing configurations in the transition region for $B=3$. The external field is $H = 0.10$.}
     \label{fig:B3_configurations}
\end{figure}

\begin{figure}[h!]
  \centering
  \includegraphics[width=\linewidth]{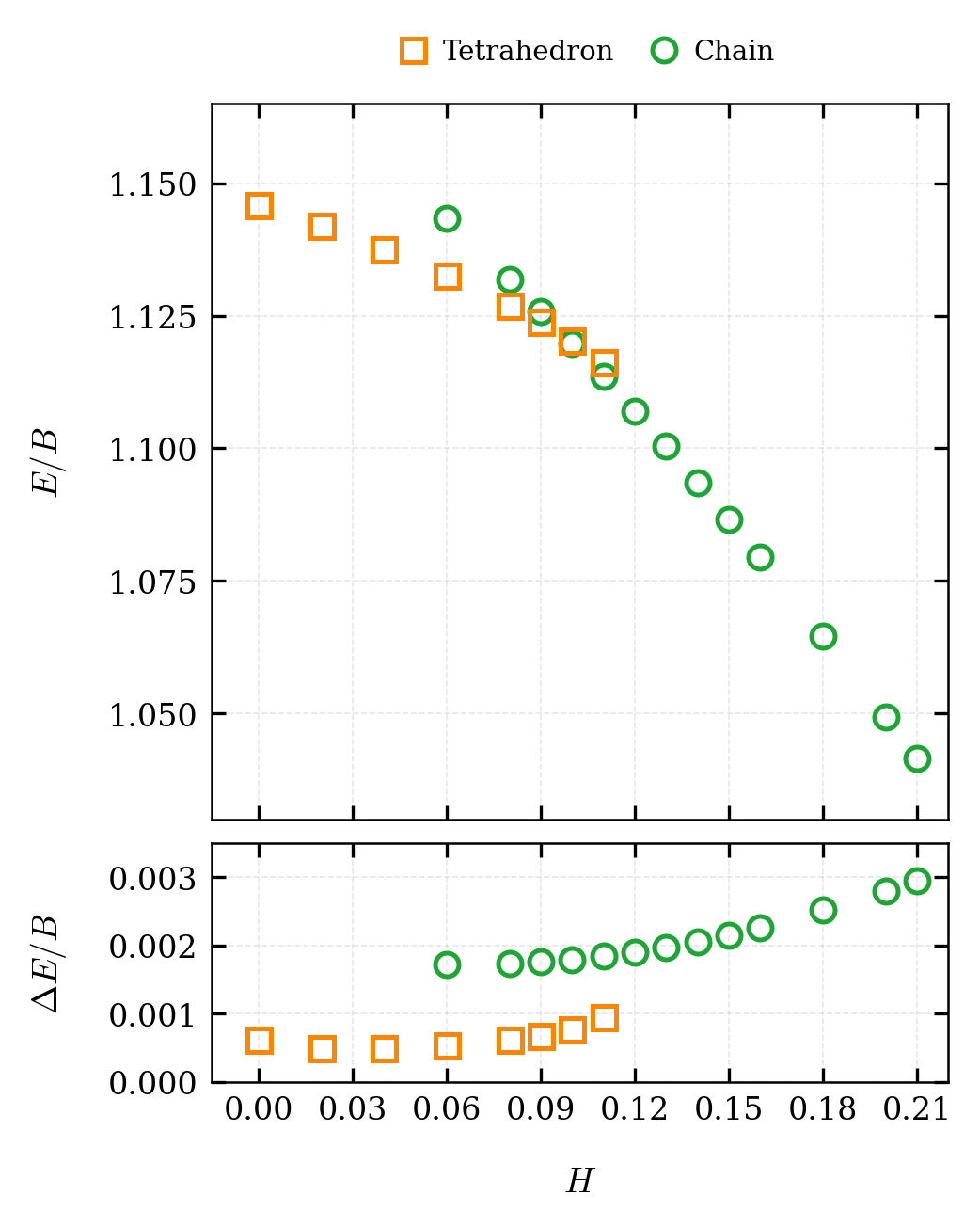}
  \caption{Energy per baryon unit for $B=3$ in the weak field regime.}
  \label{fig:B3_energy}
\end{figure}

For $B=3$, two configurations compete for the global energy minimum at weak field: a tetrahedral configuration and a $2$-toroidal chain. Their baryon density isosurfaces are shown in \cref{fig:B3_configurations}. In the chain configuration, the two holes lie in the same plane of constant $z$. The matching colour patterns on the two isosurfaces indicate that the tetrahedron can be viewed as a bent version of the chain. \Cref{fig:B3_energy} shows how their energies depend on the induction. The tetrahedron has lower energy for $H\leq0.09$, and the two curves cross at $H=0.10$. For $H\in[0.10,0.21]$ the chain is the global minimizer, and for $H\geq0.22$ it is replaced by the toroid, as summarized in \cref{fig:minimizer_intervals}.

\begin{figure}[h!]
    \centering
    \begin{subfigure}[b]{0.32\linewidth}
        \centering
        \includegraphics[width=\linewidth]{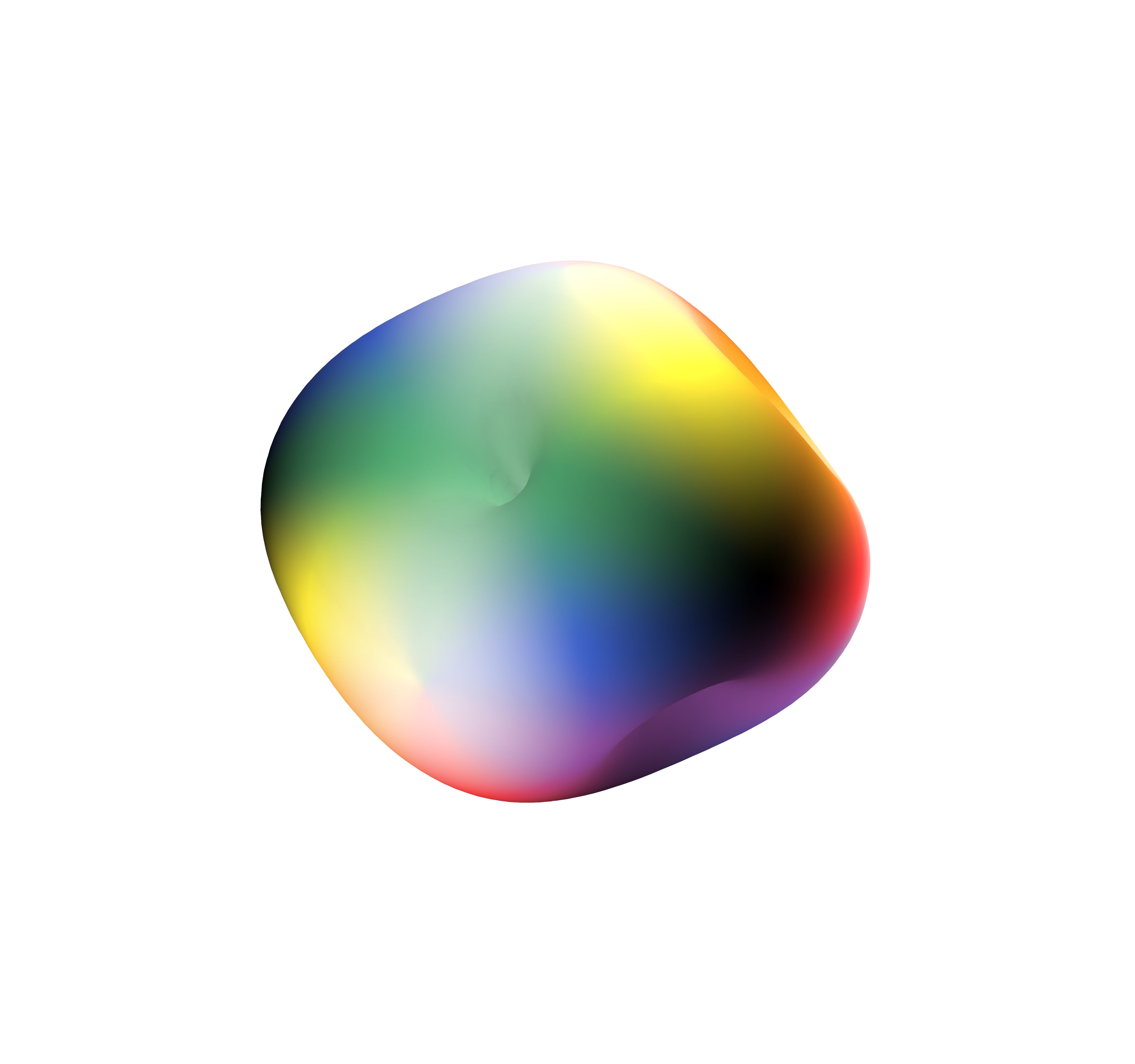}
        \caption{\begin{tabular}[t]{@{}l@{}}Rhombo-\\ hedron\end{tabular}}
    \end{subfigure}
    \hfill
    \begin{subfigure}[b]{0.32\linewidth}
        \centering
        \includegraphics[width=\linewidth]{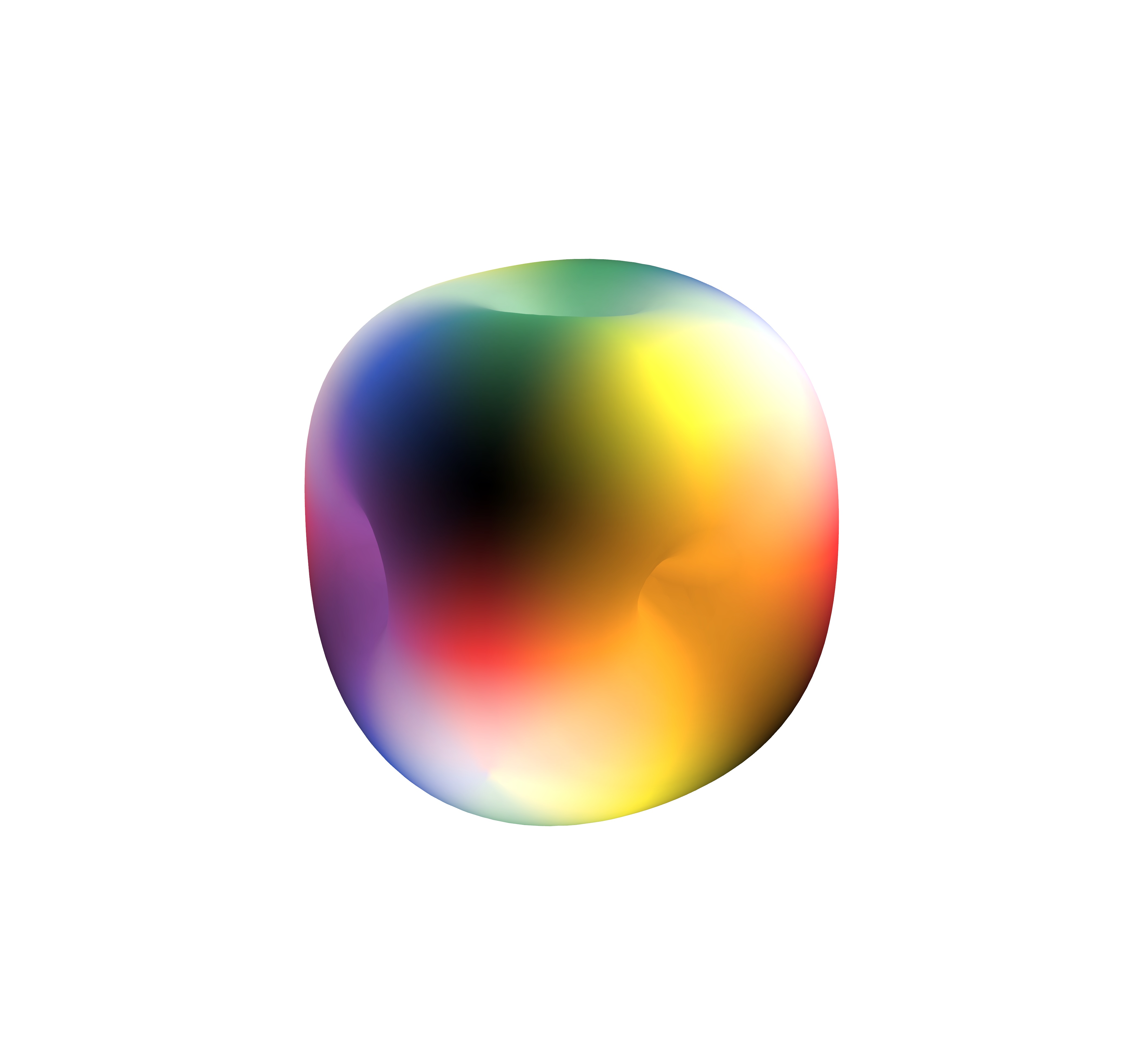}
        \caption{\begin{tabular}[t]{@{}l@{}}Rectangular\\ prism\end{tabular}}
    \end{subfigure}
    \hfill
    \begin{subfigure}[b]{0.32\linewidth}
        \centering
        \includegraphics[width=\linewidth]{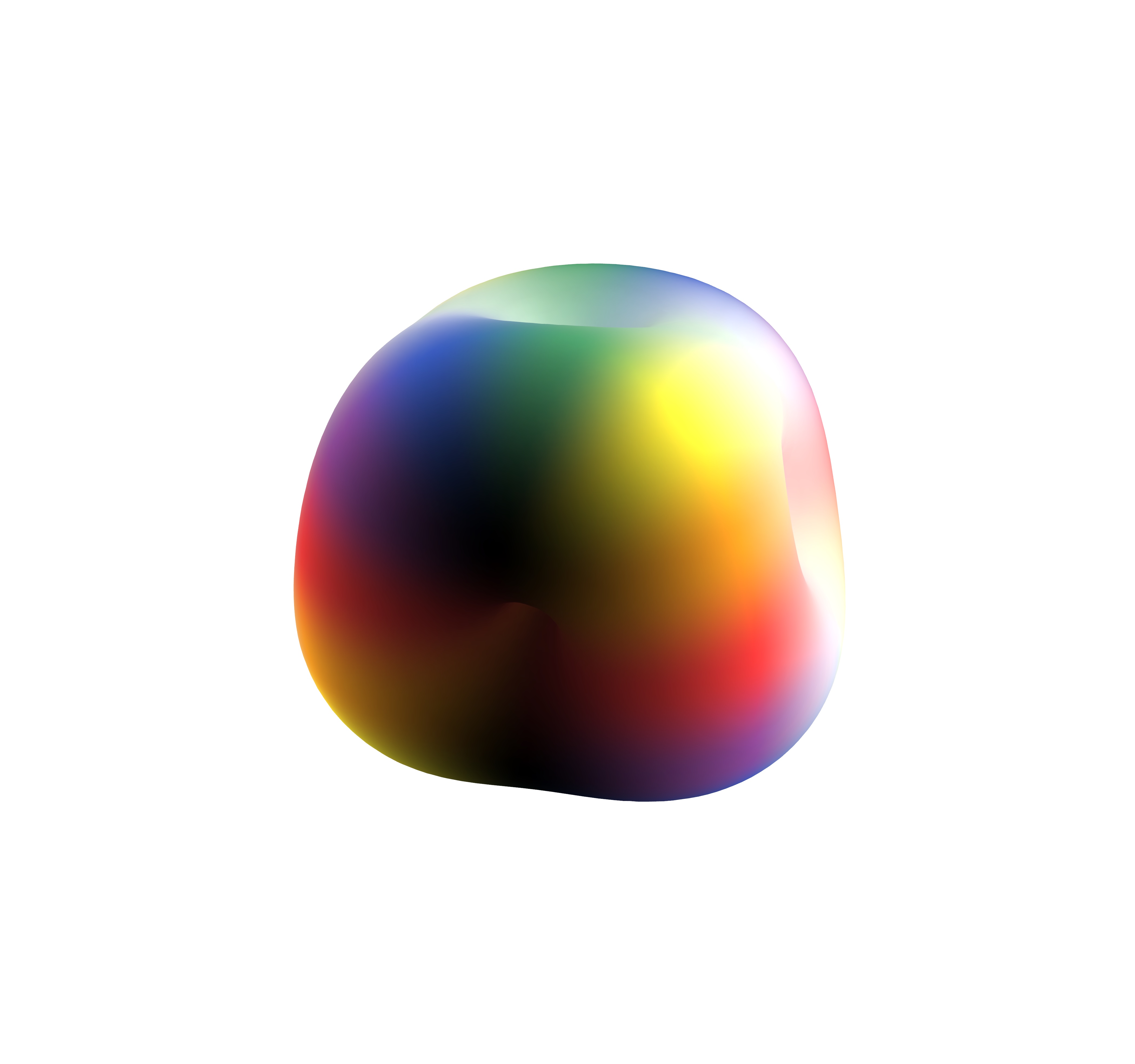}
        \caption{\begin{tabular}[t]{@{}l@{}}Molar\\ ~\end{tabular}}
    \end{subfigure}

    \medskip

    \begin{subfigure}[b]{0.32\linewidth}
        \centering
        \includegraphics[width=\linewidth]{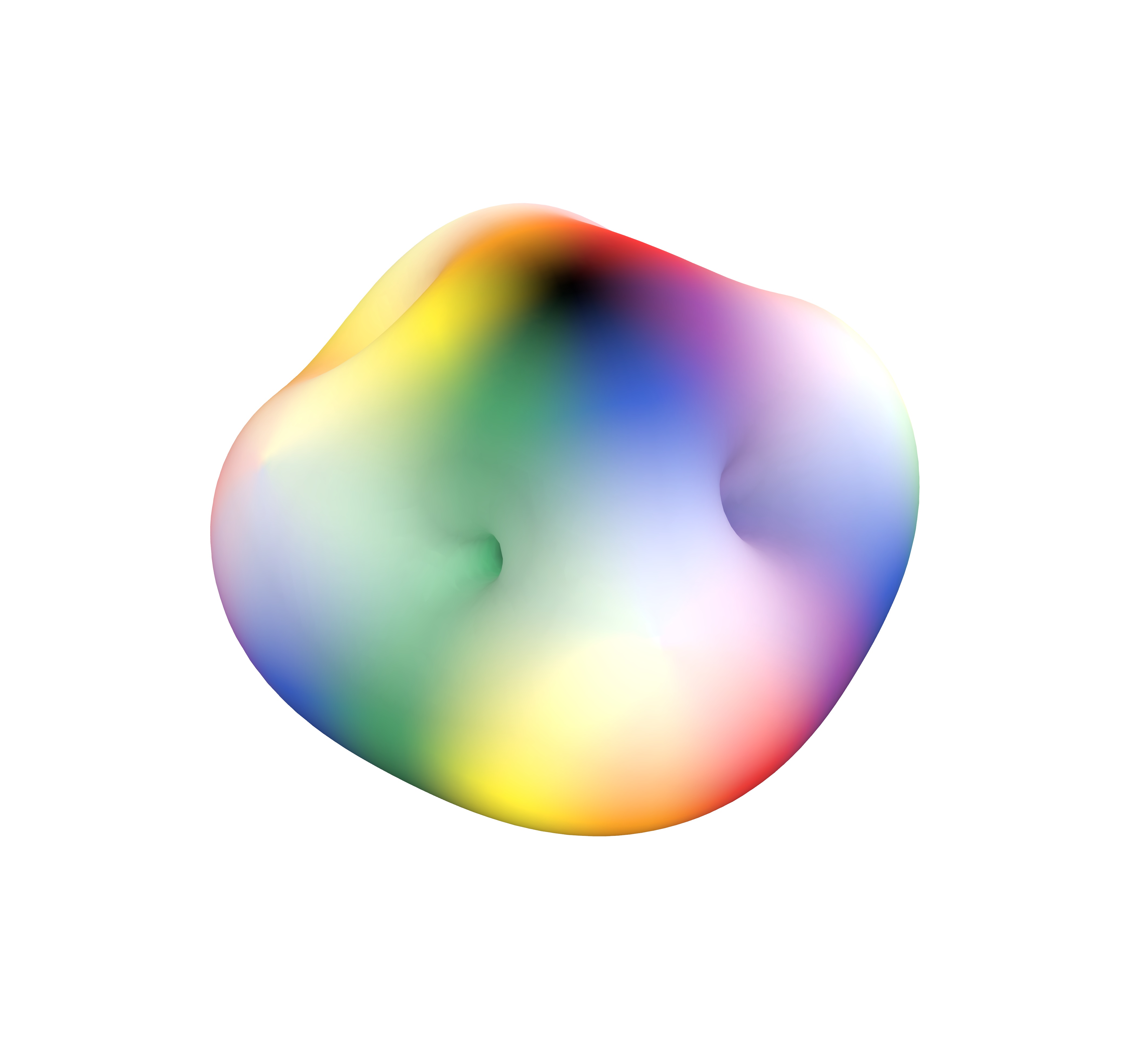}
        \caption{\begin{tabular}[t]{@{}l@{}}Square\\ pyramid\end{tabular}}
    \end{subfigure}
    \hfill
    \begin{subfigure}[b]{0.32\linewidth}
        \centering
        \includegraphics[width=\linewidth]{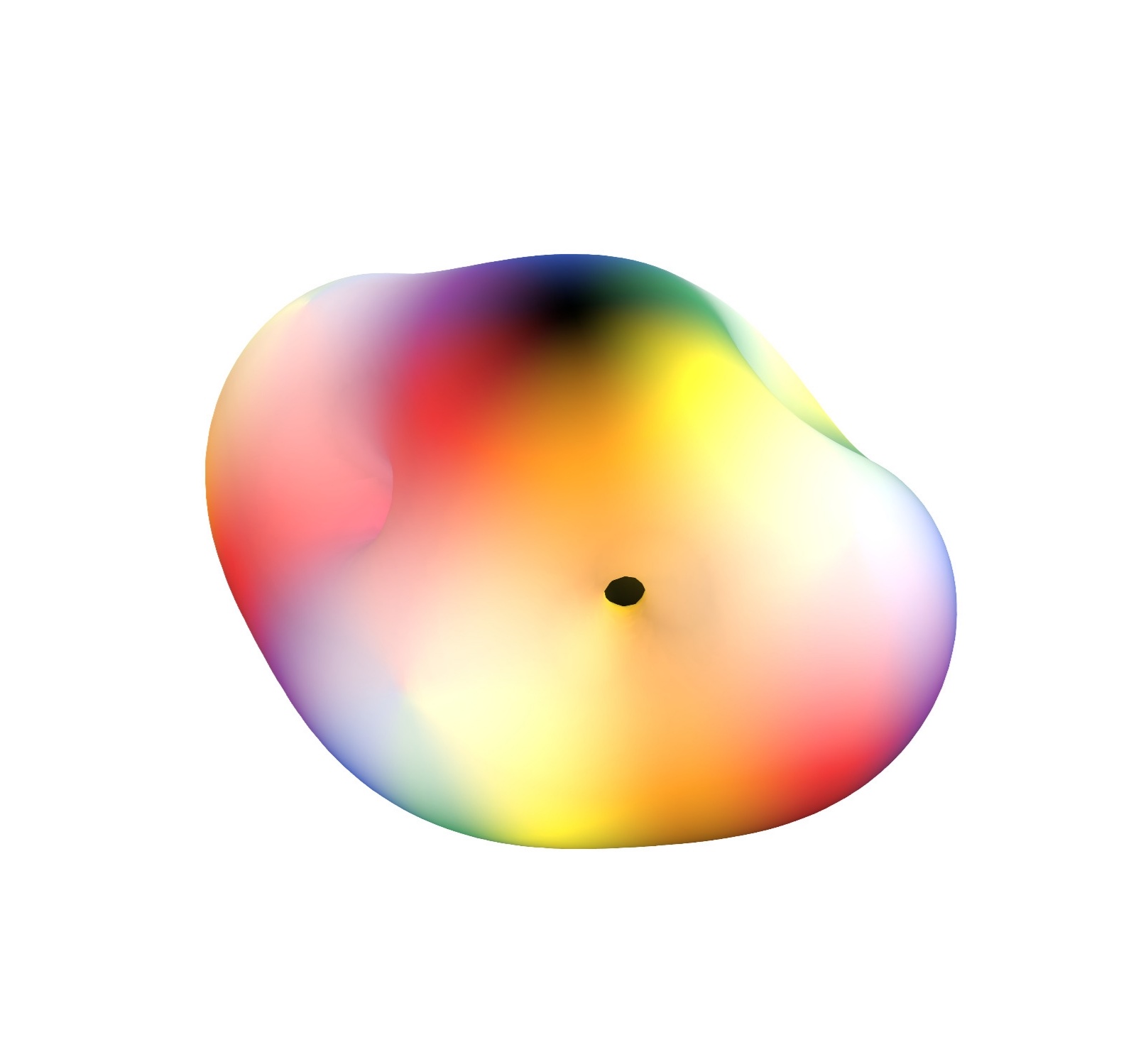}
        \caption{\begin{tabular}[t]{@{}l@{}}Rhombic\\ pyramid\end{tabular}}
    \end{subfigure}
    \hfill
    \begin{subfigure}[b]{0.32\linewidth}
        \centering
        \includegraphics[width=\linewidth]{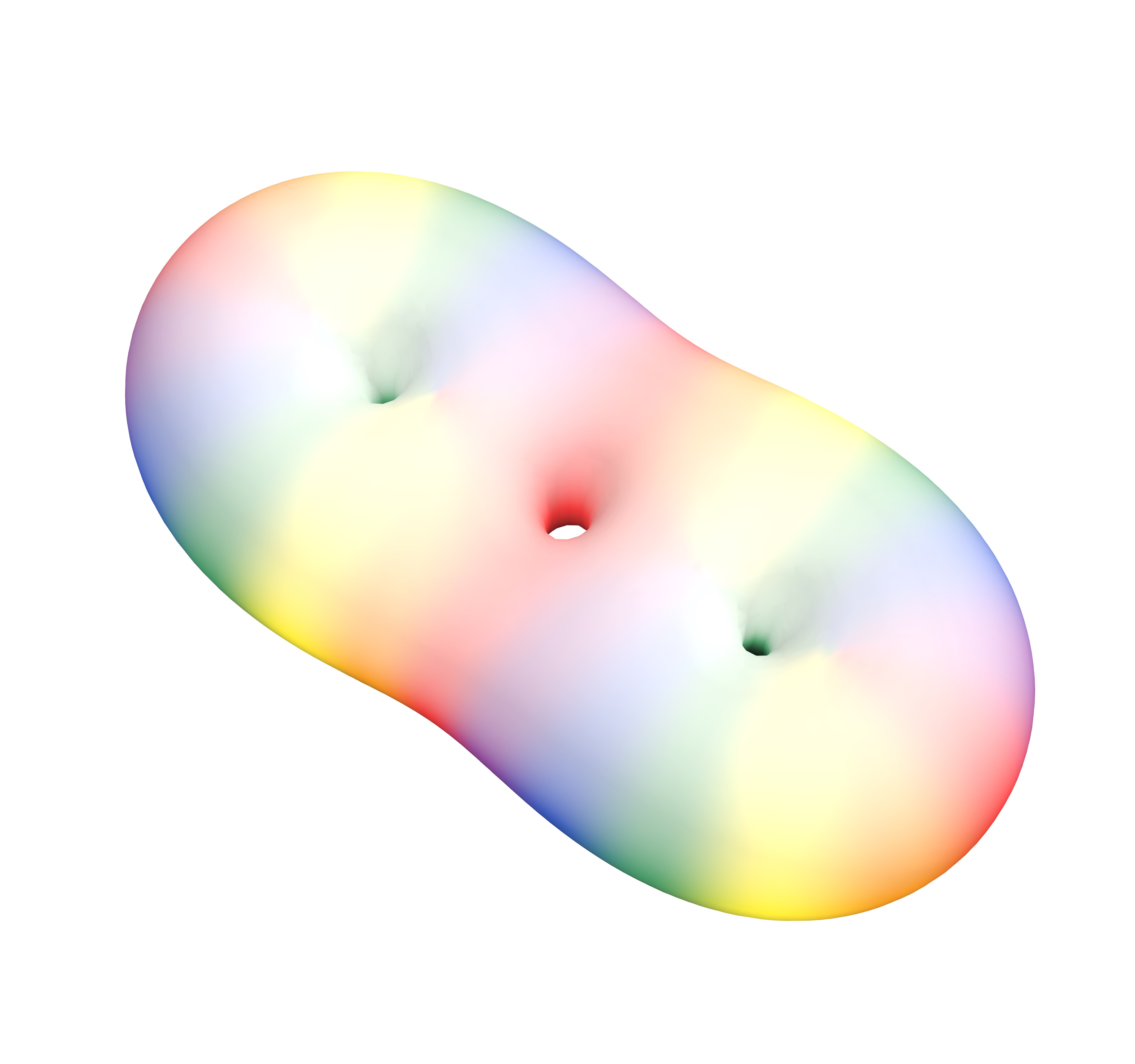}
        \caption{\begin{tabular}[t]{@{}l@{}}Chain\\ ~\end{tabular}}
    \end{subfigure}

    \caption{Competing configurations in the transition region for $B=4$. The external field is $H = 0.10$.}
    \label{fig:B4_configurations}
\end{figure}

Now we focus on the $B=4$ case. At weak magnetic fields, the configuration space is very flat, with up to six configurations having comparable energy. Three of them are deformations of the cubic rational-map shape: a \textit{rectangular prism}, a \textit{rhombohedron}, and a shape with trapezoidal side faces which we call the \textit{molar}. The remaining three are a \textit{square pyramid}, a \textit{rhombic pyramid} and a $3$-toroidal \textit{chain}, analogous to the $B=3$ chain. All these baryon density geometries are shown in \cref{fig:B4_configurations}.

\begin{figure}[h!]
  \centering
  \includegraphics[width=\linewidth]{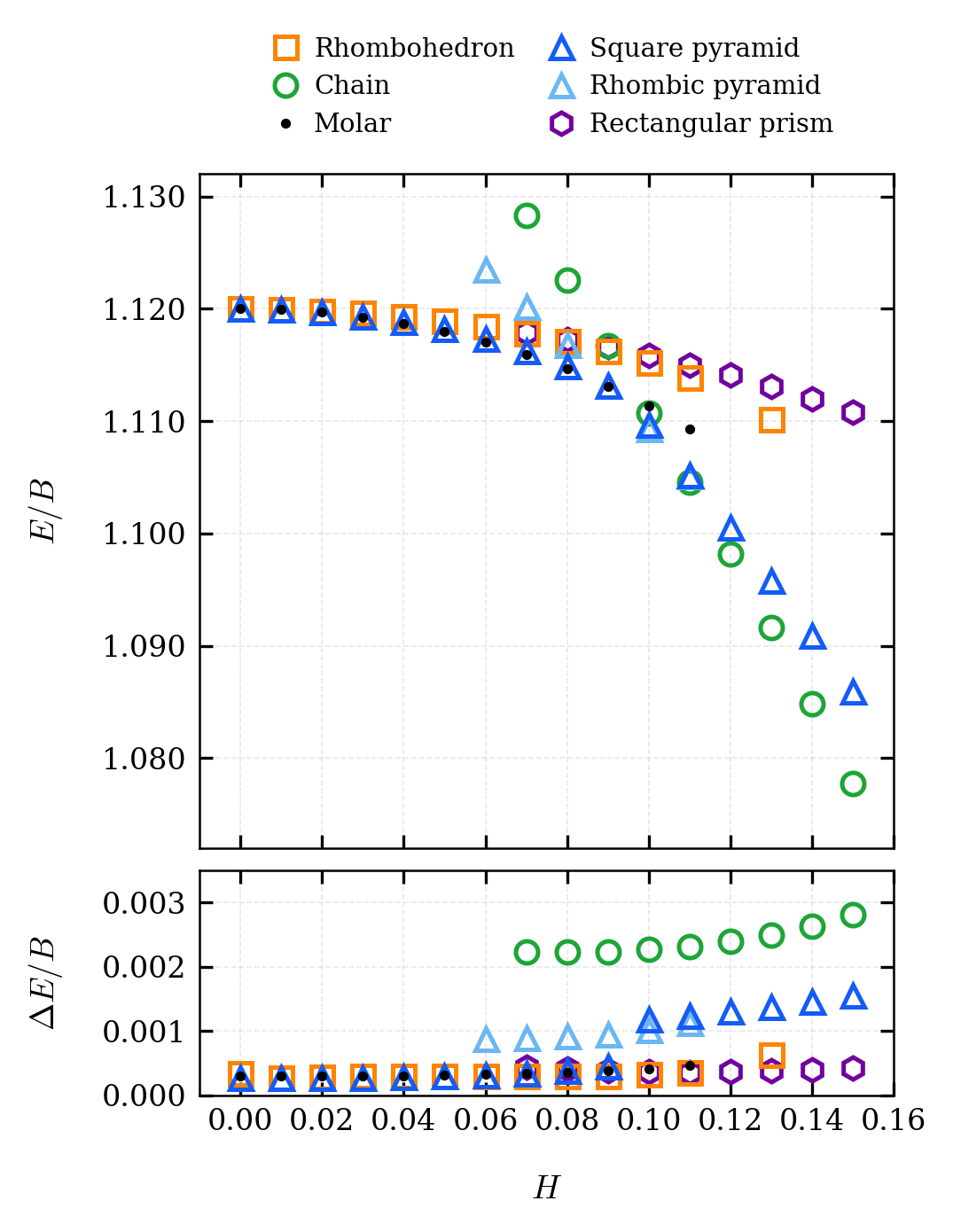}
  \caption{Energy per baryon unit for $B=4$ in the weak field regime.}
  \label{fig:B4_energy}
\end{figure}

The energy of the competing configurations is shown in \cref{fig:B4_energy}. At $H=0.00$ the rhombohedron, the molar and the square pyramid all relax to the same cubic shape. For $H<0.09$ the molar has the lowest energy, but the difference with the square pyramid is within the numerical error, as is the difference with the rhombohedron for $H<0.05$. At $H=0.10$ the rhombic pyramid has the lowest energy, again within error of the square pyramid and the chain. For $0.10<H<0.20$ the chain is clearly the minimizer, while for $H>0.20$ it is replaced by the toroid, as indicated in \cref{fig:minimizer_intervals}.

\newpage
\subsubsection{Strong magnetic field chains and toroids}
\begin{figure}[h!]
    \centering

    \begin{subfigure}[b]{0.32\linewidth}
        \centering
        \includegraphics[width=\linewidth]{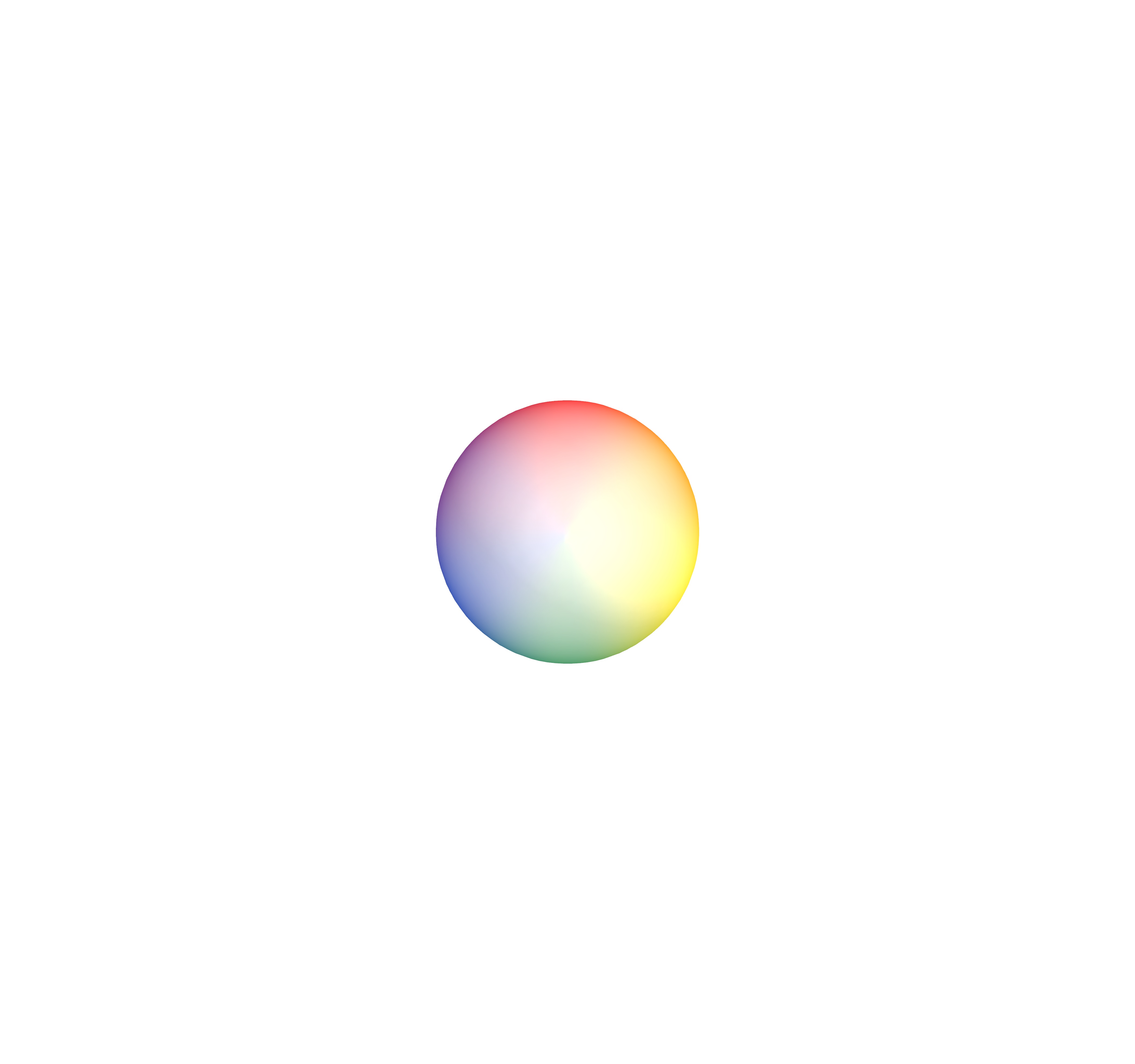}
        \caption{$B=1$}
    \end{subfigure}
    \hfill
    \begin{subfigure}[b]{0.32\linewidth}
        \centering
        \includegraphics[width=\linewidth]{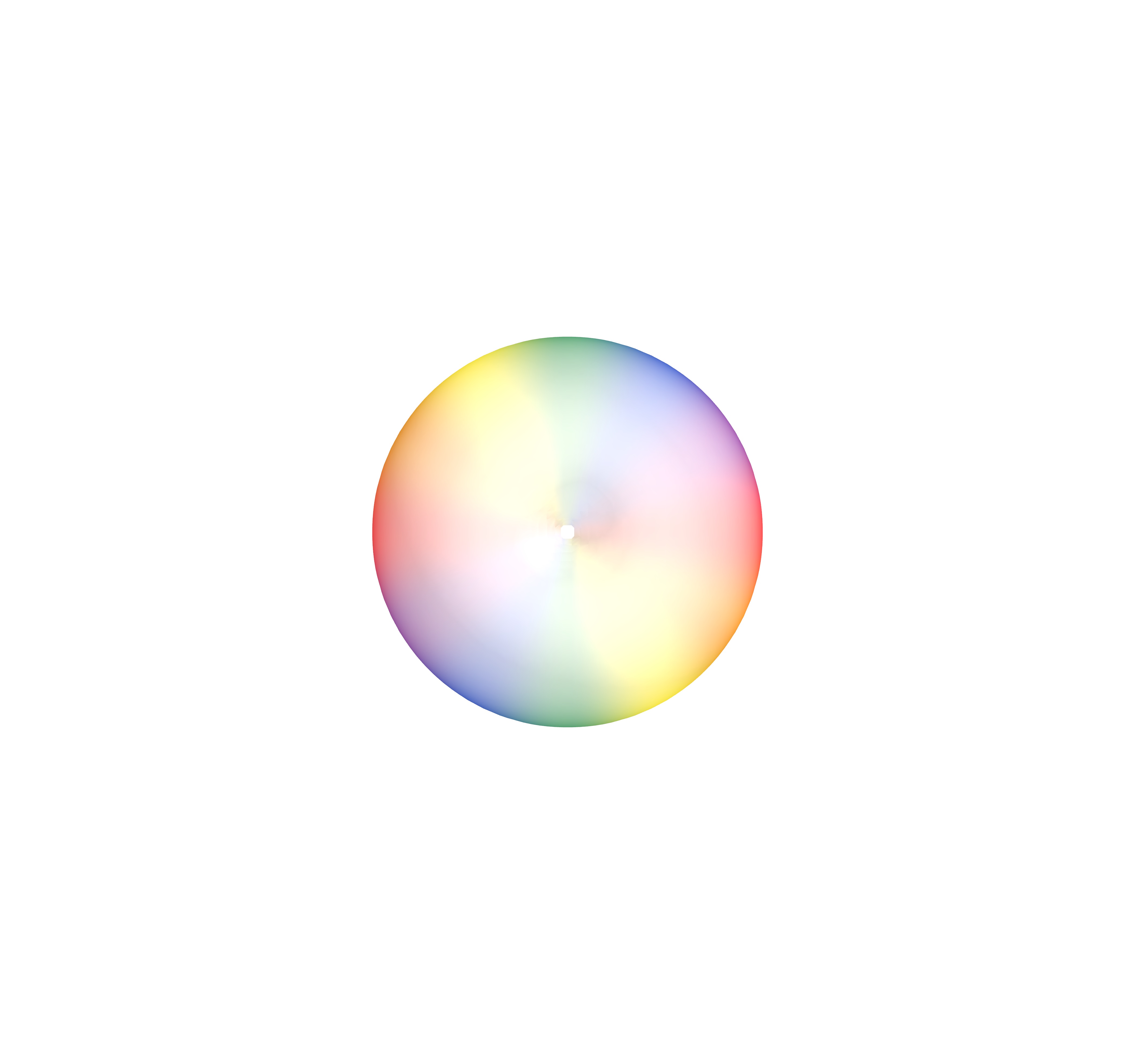}
        \caption{$B=2$}
    \end{subfigure}
    \hfill
    \begin{subfigure}[b]{0.32\linewidth}
        \centering
        \includegraphics[width=\linewidth]{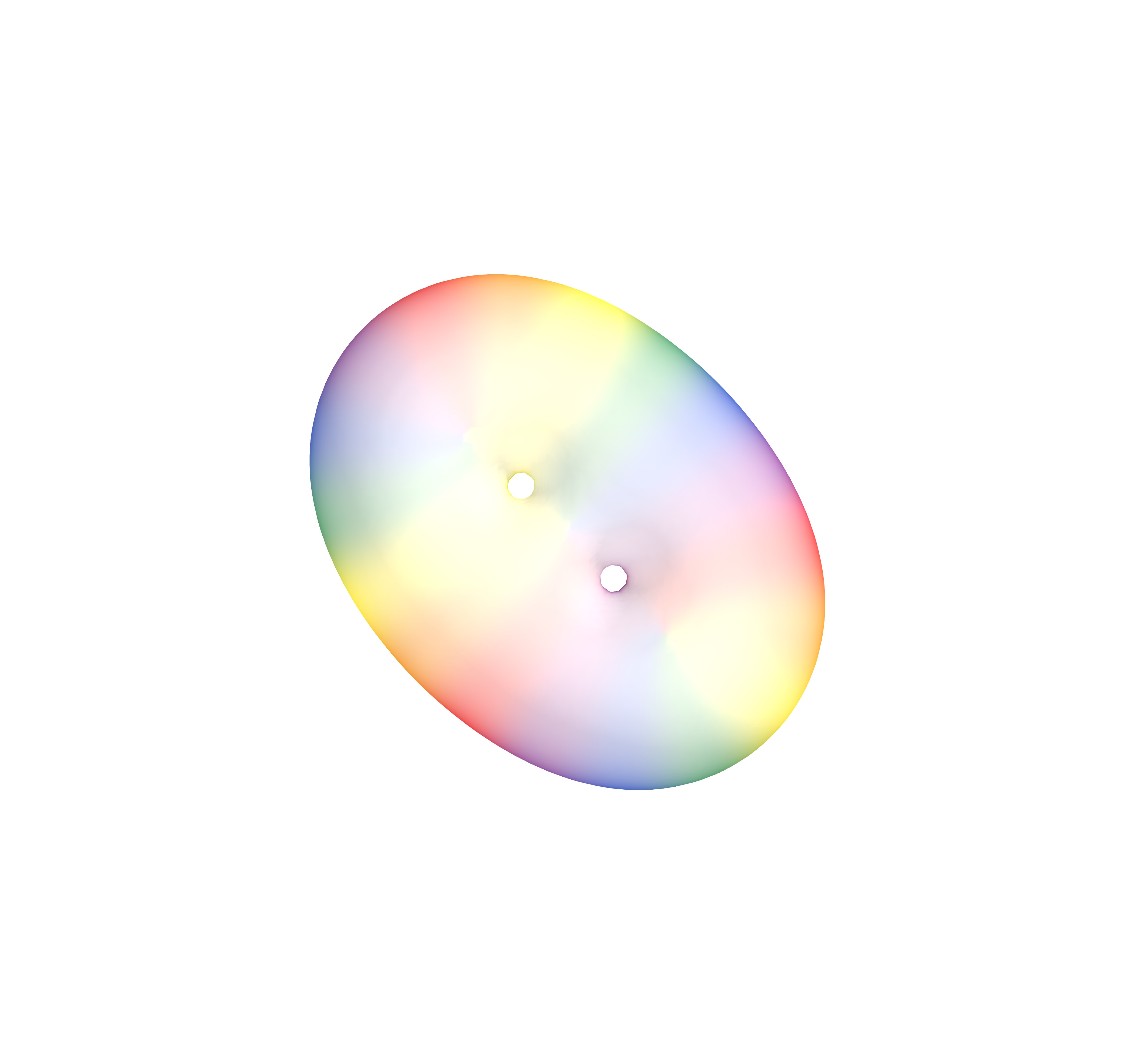}
        \caption{$B=3$}
    \end{subfigure}

    \medskip

    \begin{subfigure}[b]{0.32\linewidth}
        \centering
        \includegraphics[width=\linewidth]{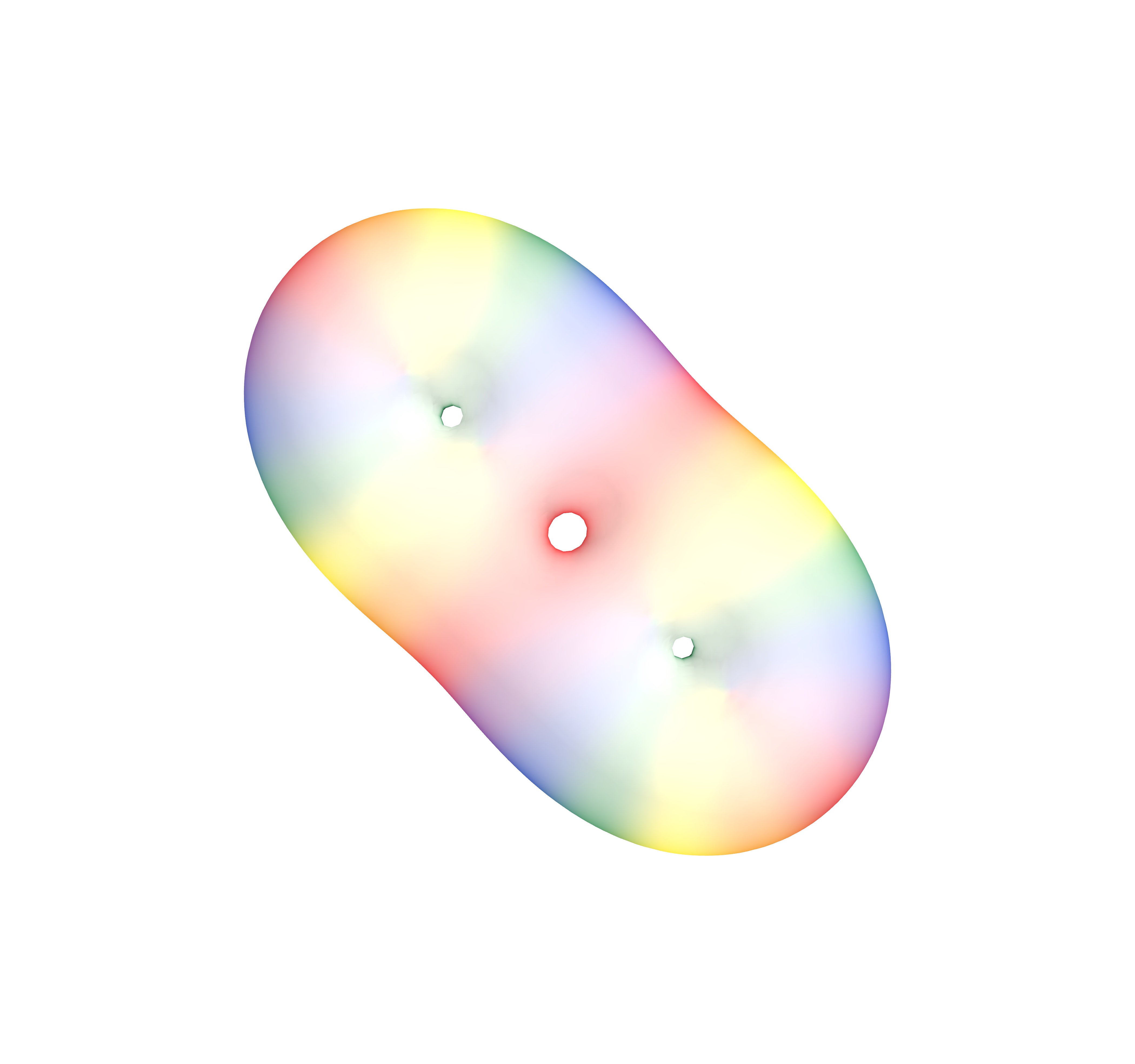}
        \caption{$B=4$}
    \end{subfigure}
    \hfill
    \begin{subfigure}[b]{0.32\linewidth}
        \centering
        \includegraphics[width=\linewidth]{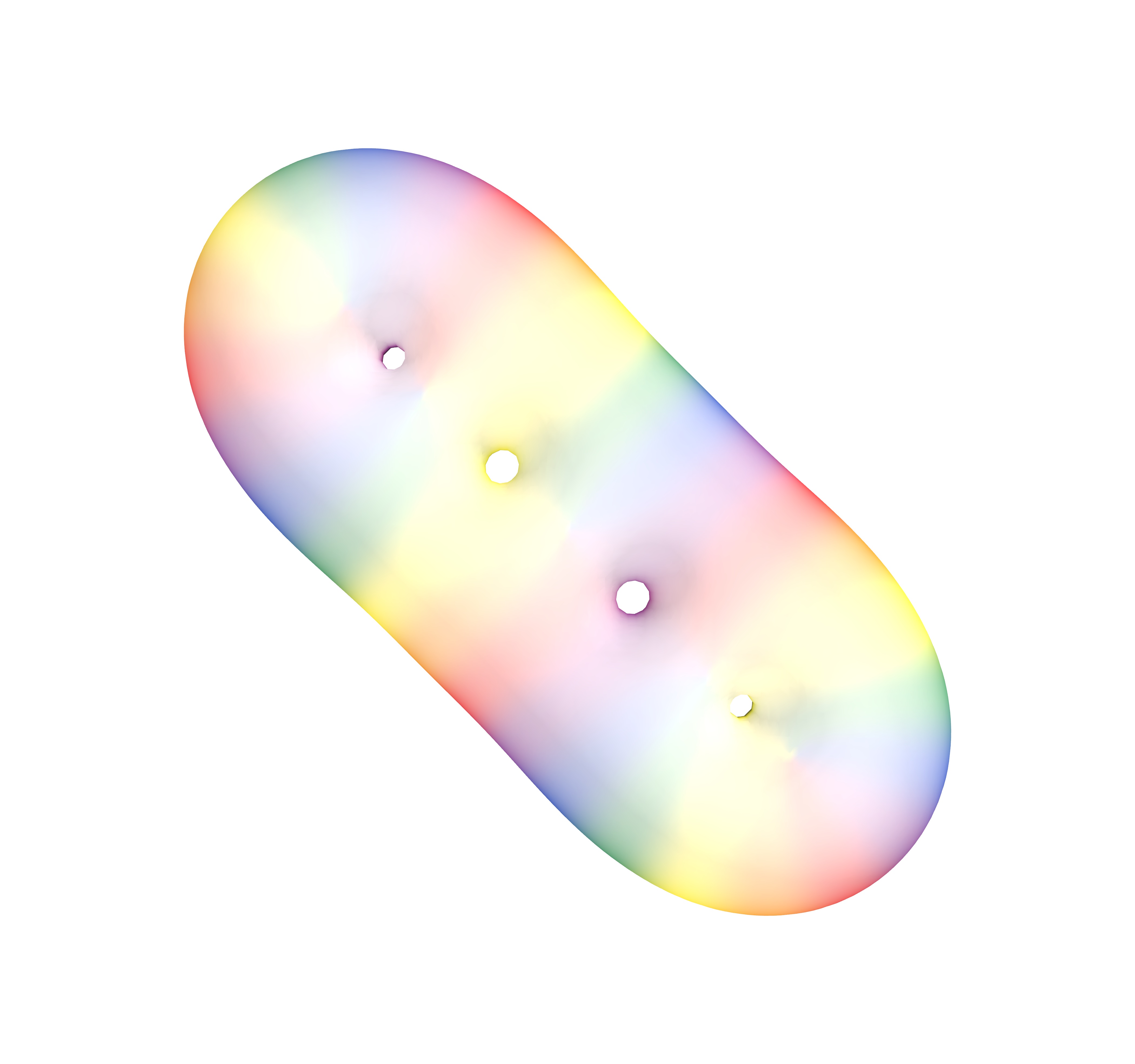}
        \caption{$B=5$}
    \end{subfigure}
    \hfill
    \begin{subfigure}[b]{0.32\linewidth}
        \centering
        \includegraphics[width=\linewidth]{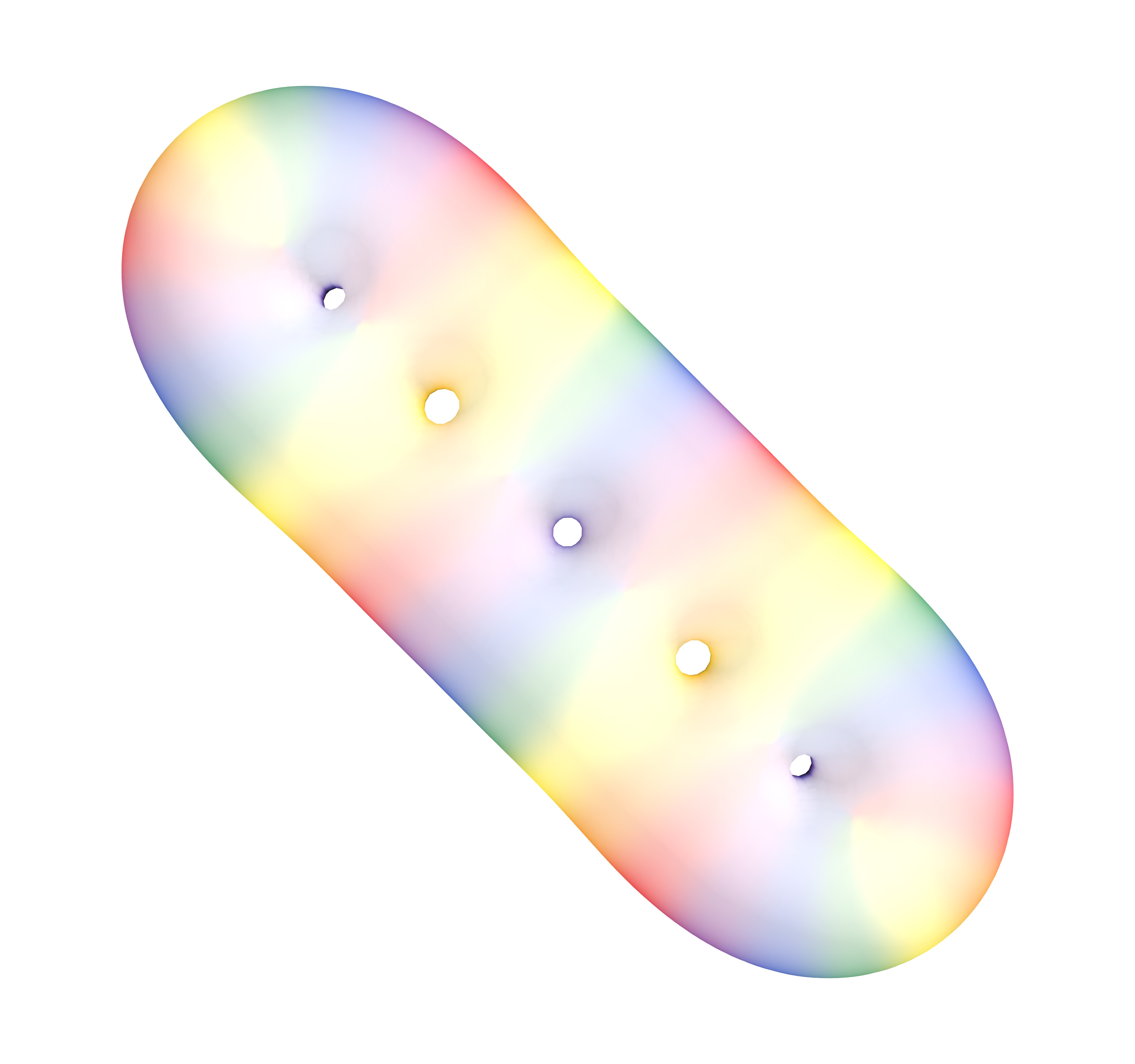}
        \caption{$B=6$}
    \end{subfigure}
    \caption{An intermediate field value $H=0.15$ produces $(B-1)$-toroidal chains as minimal energy configurations.}
    \label{fig:chains}
\end{figure}

The chain configurations found for $B=3$ and $B=4$ represent a general feature of the model. Over a narrow range of $H$, the baryon density isosurfaces across all $B$ sectors resemble a $g$-toroidal shape with $g=B-1$. The $g$ holes lie in a plane of constant $z$ and are aligned along a straight line, as illustrated in \cref{fig:chains}. The magnetic field intervals in which the chain acts as the global minimizer depends on the baryon number. They become narrower as $B$ increases, as indicated in \cref{fig:minimizer_intervals}. Similar chains structures have been reported in the $(2+1)$-dimensional Baby Skyrme model \cite{winyard2016skyrme,Foster2010}.

\begin{figure}[H]
    \centering

    \begin{subfigure}[b]{0.32\linewidth}
        \centering
        \includegraphics[width=\linewidth]{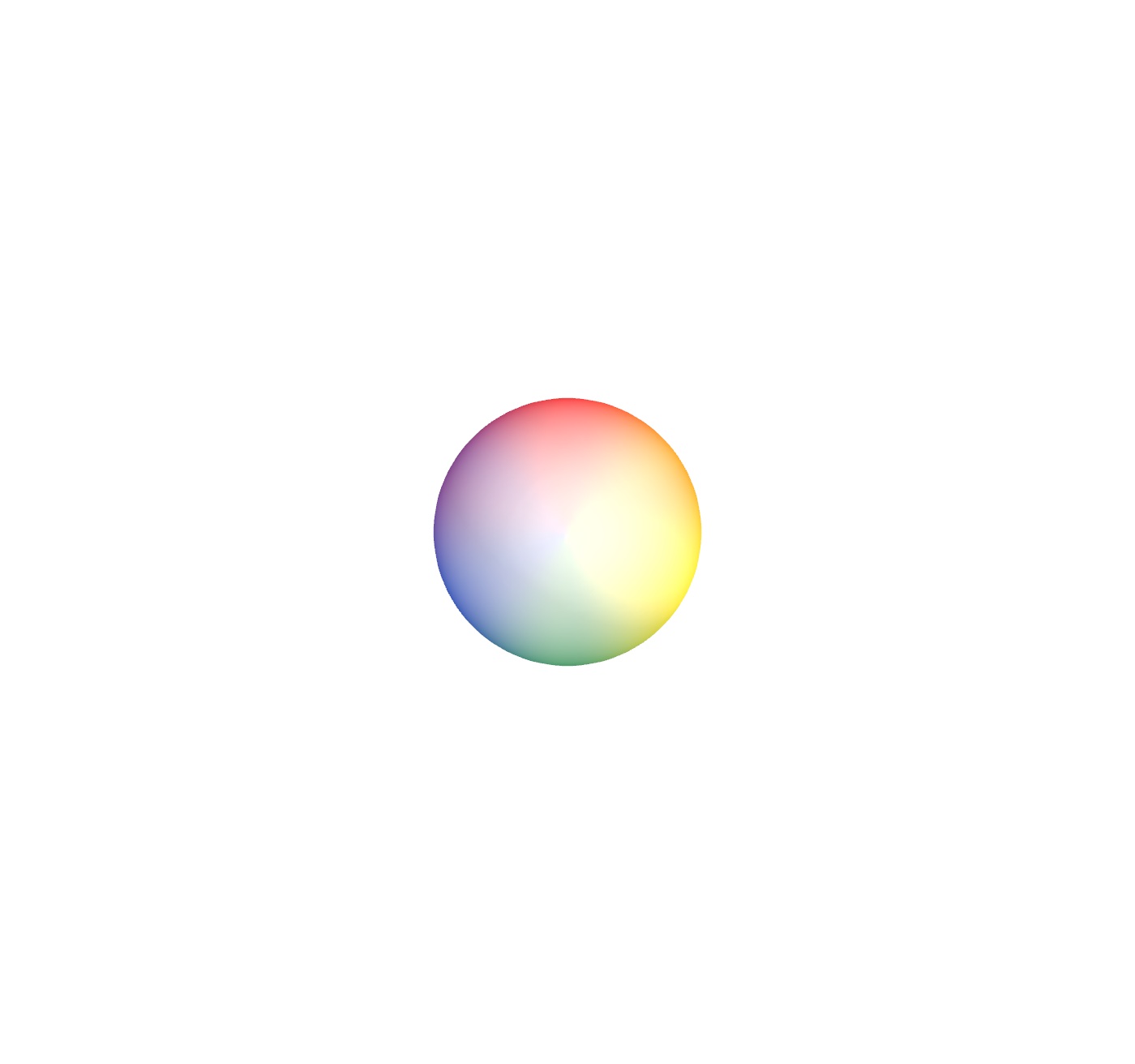}
        \caption{$B=1$}
    \end{subfigure}
    \hfill
    \begin{subfigure}[b]{0.32\linewidth}
        \centering
        \includegraphics[width=\linewidth]{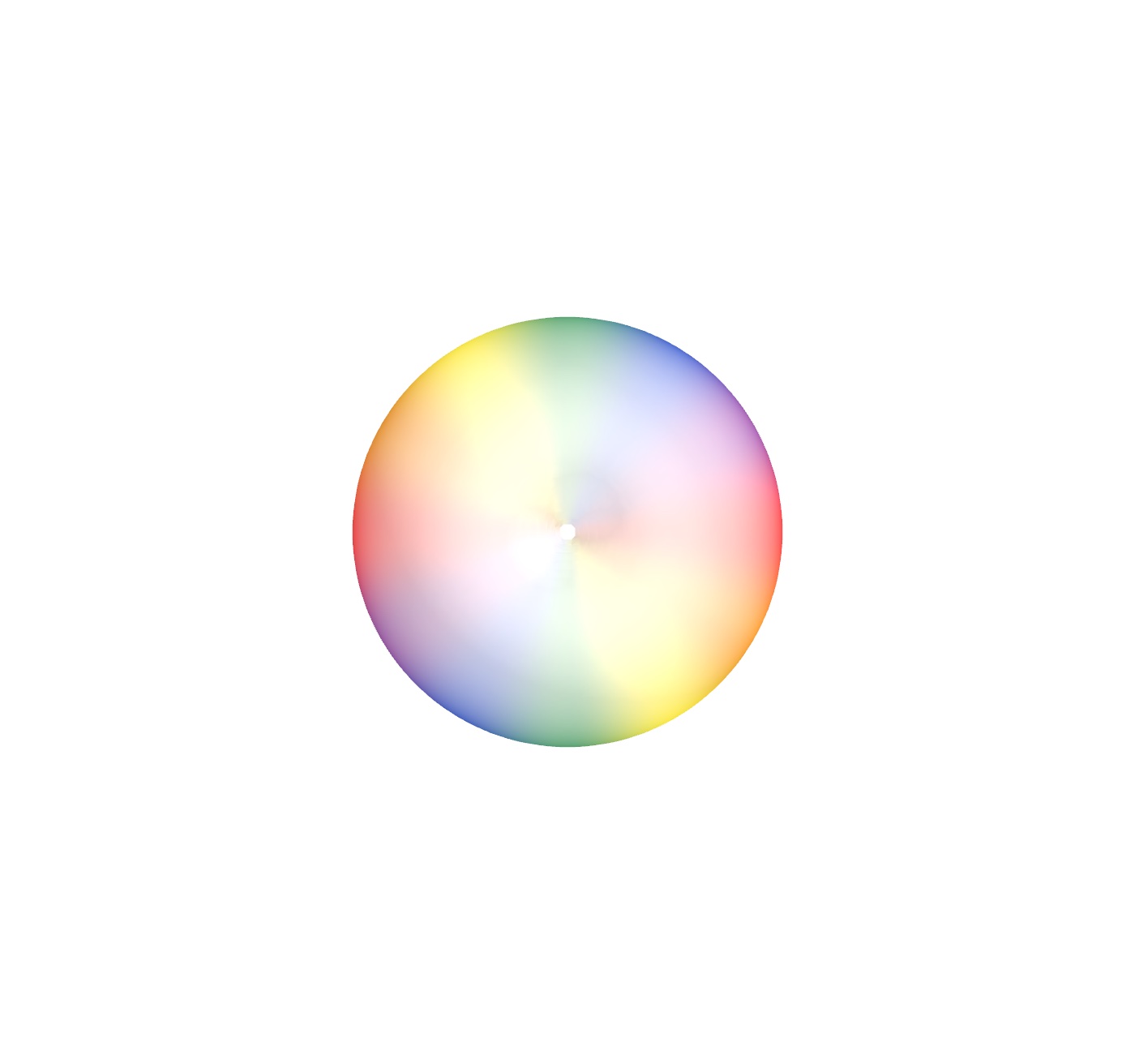}
        \caption{$B=2$}
    \end{subfigure}
    \hfill
    \begin{subfigure}[b]{0.32\linewidth}
        \centering
        \includegraphics[width=\linewidth]{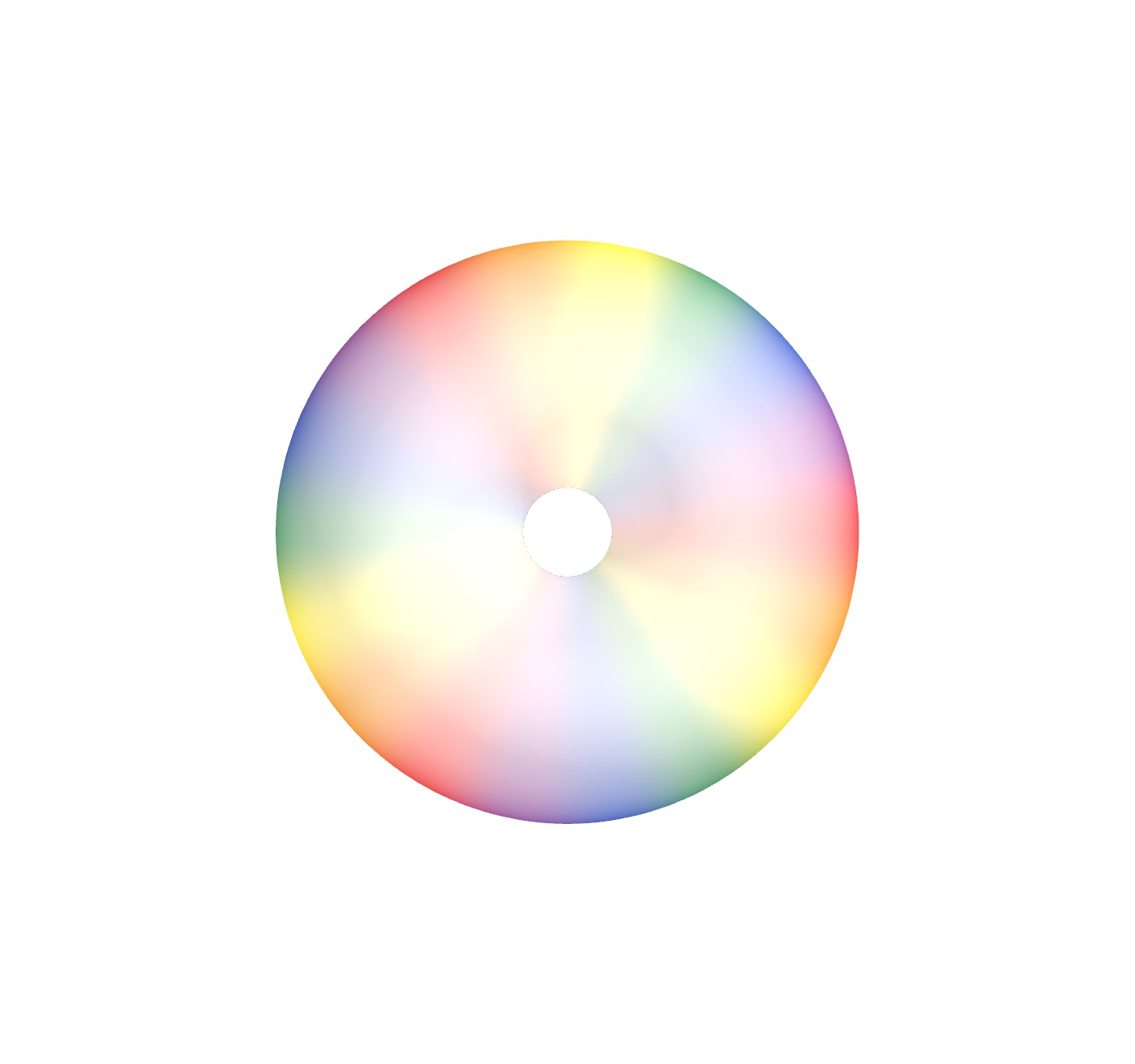}
        \caption{$B=3$}
    \end{subfigure}

    \medskip

    \begin{subfigure}[b]{0.32\linewidth}
        \centering
        \includegraphics[width=\linewidth]{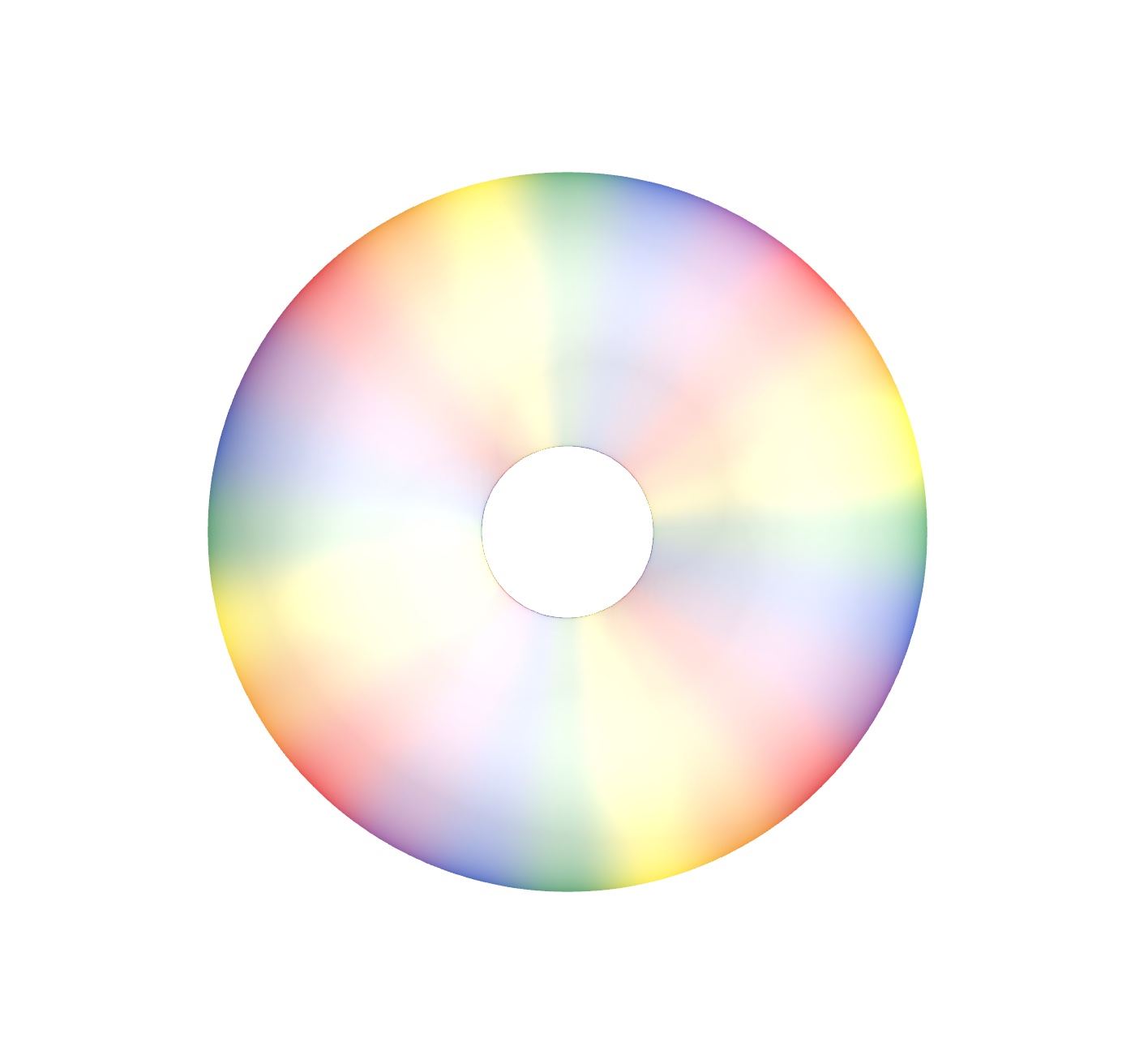}
        \caption{$B=4$}
    \end{subfigure}
    \hfill
    \begin{subfigure}[b]{0.32\linewidth}
        \centering
        \includegraphics[width=\linewidth]{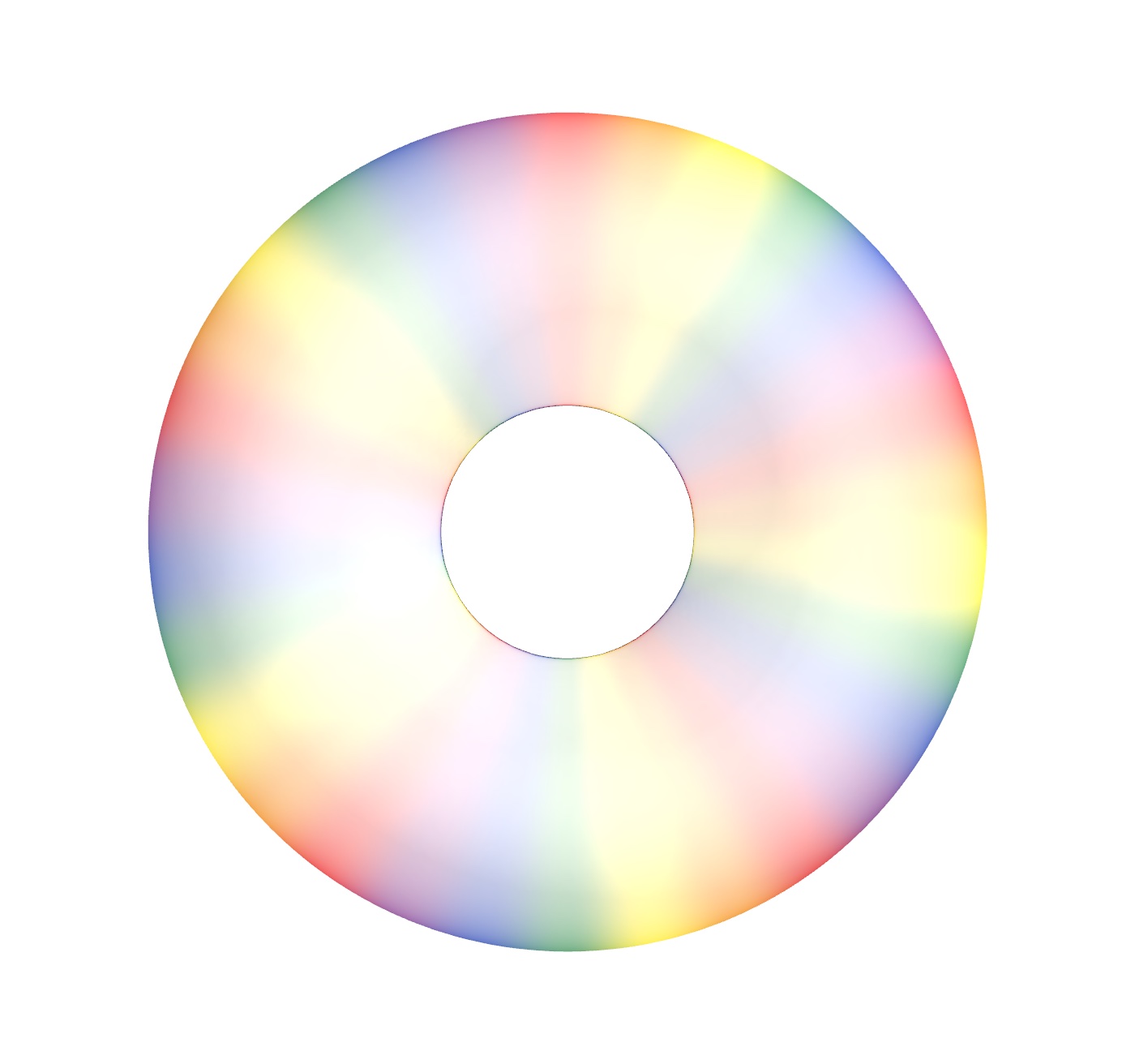}
        \caption{$B=5$}
    \end{subfigure}
    \hfill
    \begin{subfigure}[b]{0.32\linewidth}
        \centering
        \includegraphics[width=\linewidth]{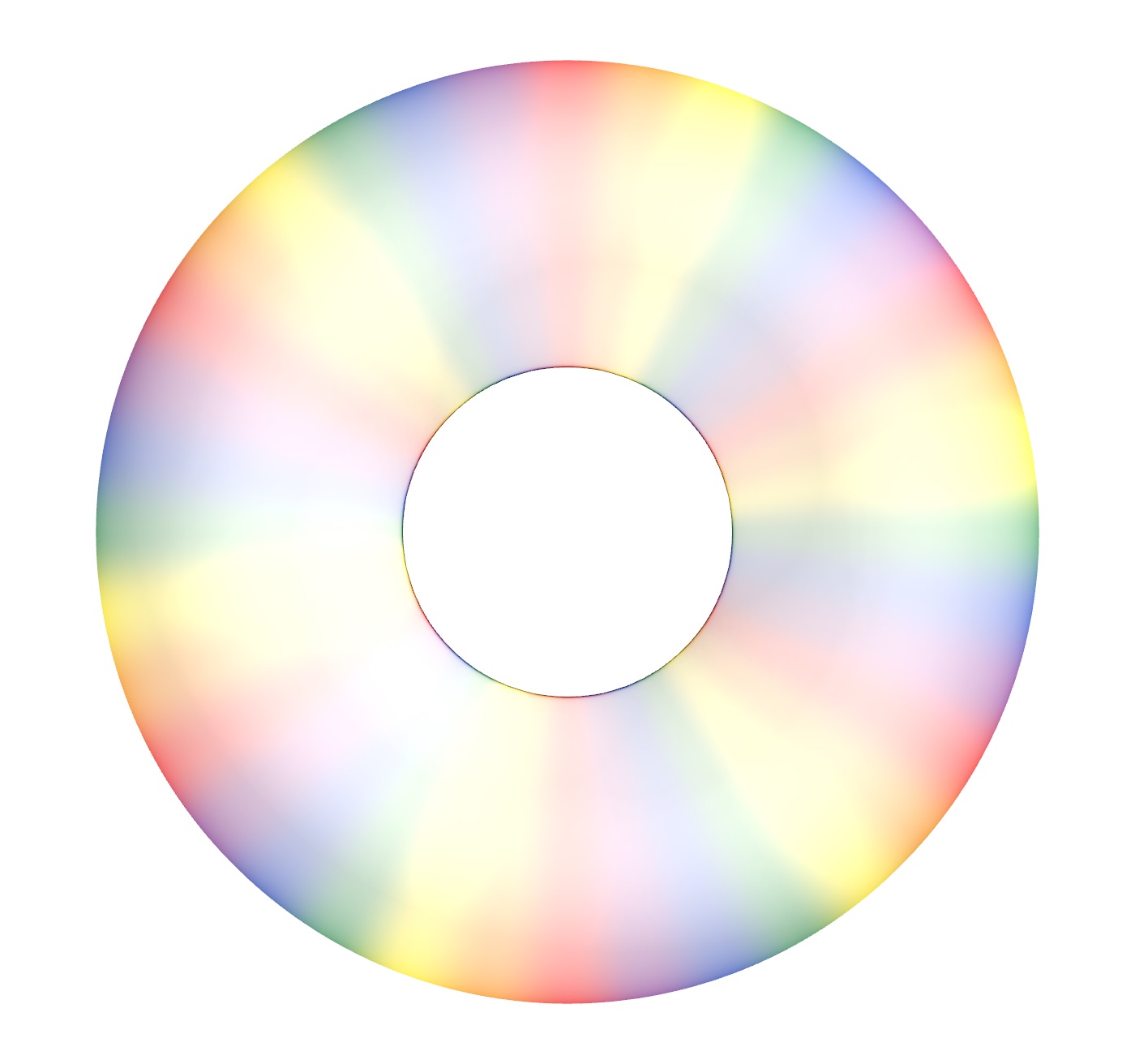}
        \caption{$B=6$}
    \end{subfigure}

    \caption{A strong field value $H=0.40$ produces toroids as minimal energy configurations.}
    \label{fig:tori}
\end{figure}

For stronger magnetic fields, $H\geq0.22$, the static configurations for all baryon numbers become axially symmetric toroids, whose symmetry axis are aligned with the field $\hat{z}$. The isosurfaces with this geometry are shown in \cref{fig:tori}. The colour pattern on each isosurface repeats $B$ times around the symmetry axis, so the configuration is invariant under rotations by an azimuthal angle $2\pi n/B$, with $n\in \mathbb{Z}$. A general rotation about this symmetry axis can be undone with a rotation in the $\phi$ space. The toroidal configurations shrink as the magnetic field strength increases, as it was illustrated in \cref{fig:all_in_one}.

\subsection{Negative baryon density regions}\label{sec:neg_density}

It is known from \cite{Houghton2001,Foster2013} that non-holomorphic rational maps give skyrmions with lower energy than the standard rational map ansatz, and closer to the numerical solutions. These maps allow folds in the skyrmion fields that locally cover $S^3$ with opposite orientation, resulting in regions of negative baryon density. The energy density, however, remains non-negative, since it contains only products of an even number of fields.

\begin{figure}[h!]
  \centering
  \begin{subfigure}{0.48\linewidth}
    \centering
    \includegraphics[width=\linewidth]{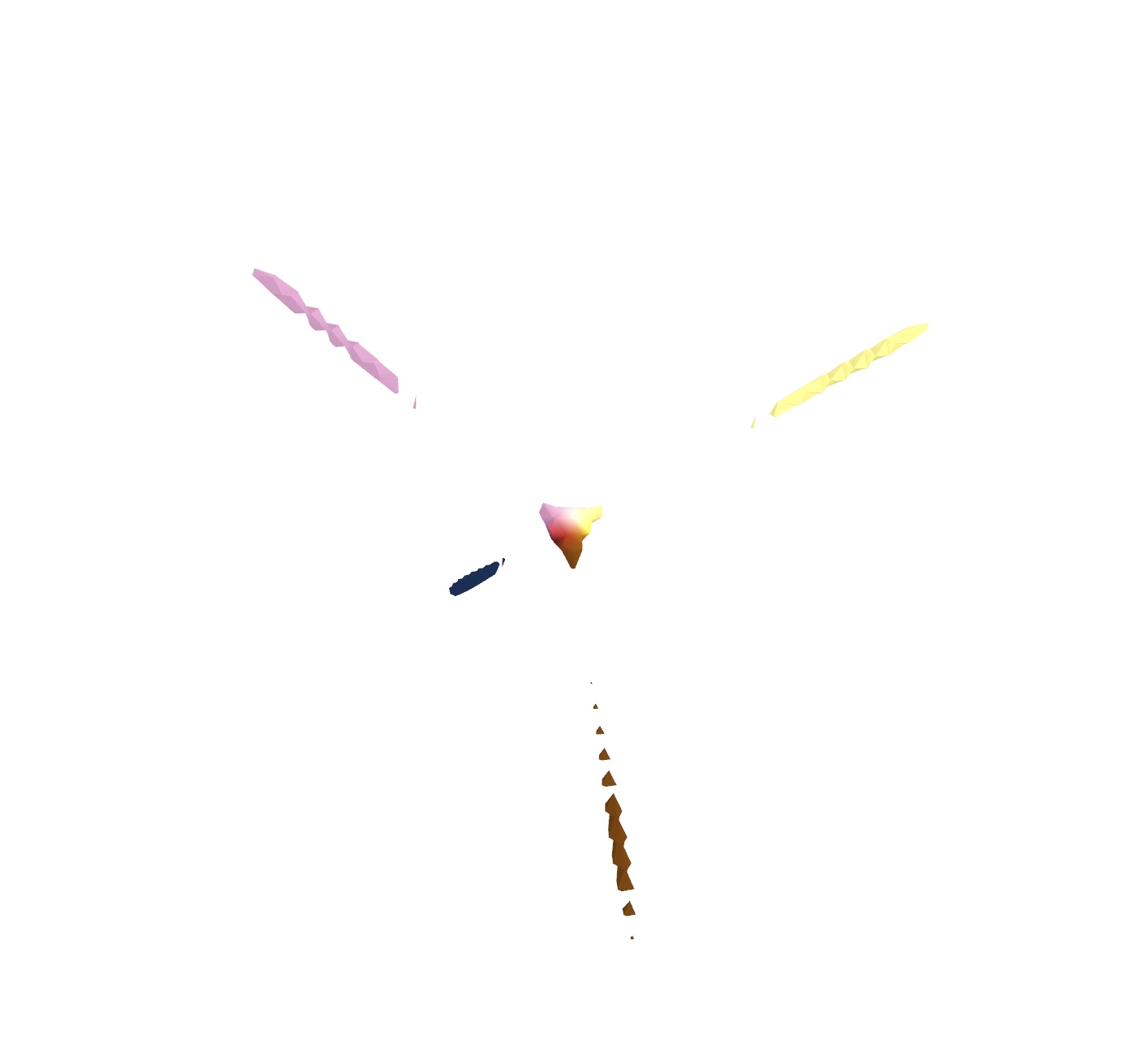}
    \caption{$B=3$}
  \end{subfigure}
  \hfill
  \begin{subfigure}{0.48\linewidth}
    \centering
    \includegraphics[width=\linewidth]{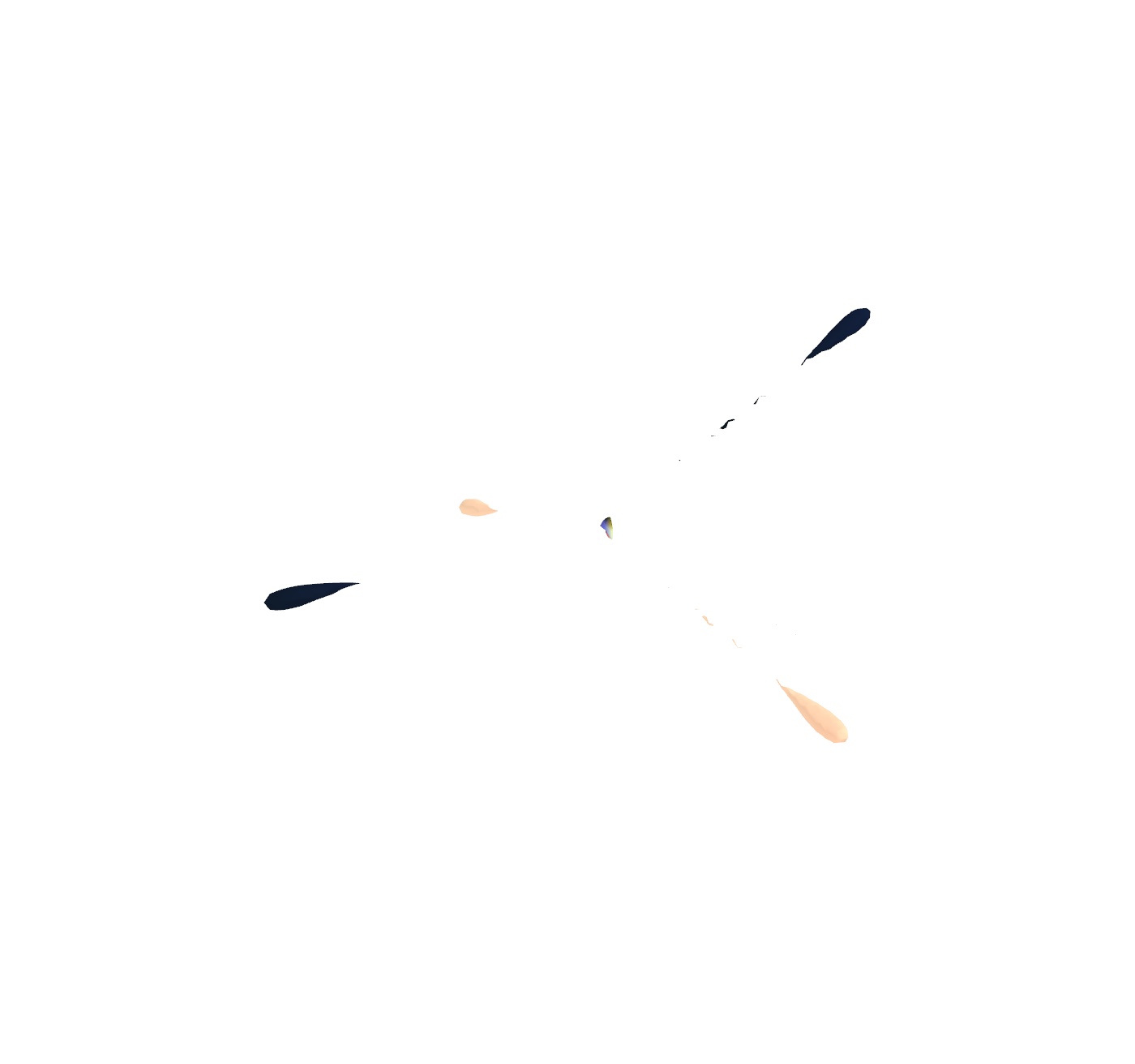}
    \caption{$B=5$}
  \end{subfigure}

  \vspace{0.5cm}

  \begin{subfigure}{0.48\linewidth}
    \centering
    \includegraphics[width=\linewidth]{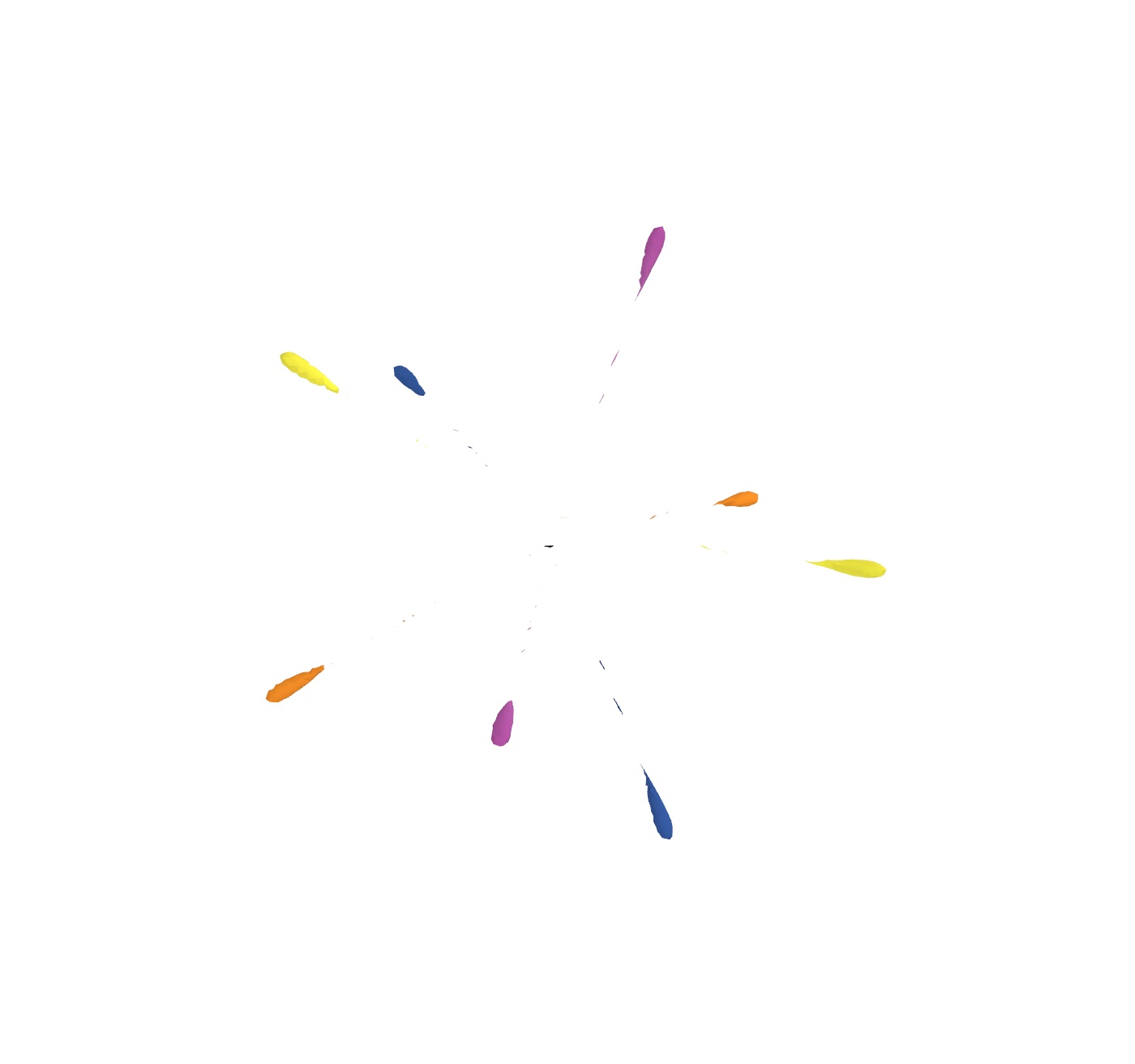}
    \caption{$B=6$}
  \end{subfigure}

  \caption{Negative baryon density regions in the ungauged case $H=0.00$.}
  \label{fig:negative_density}
\end{figure}

We find that the minimizers for $B=3,5,6$ which correspond to deformations of the rational map have folded structures featuring regions of negative baryon density. In contrast, all other configurations presented in this paper exhibit no negative baryon density at any lattice point. In all cases, the negative density forms a core around the origin from which several tubular branches extend outward, \cref{fig:negative_density}. The total negative baryon charge reaches its maximum at $H=0.00$, with the values listed for comparison in \cref{tab:neg_charge}.

\begin{table}[h!]
\centering
\begin{tabular}{c
                S[table-format=-1.2e-1]
                S[table-format=1.2e-1]}
\toprule
{$B$} & {$B_{\mathrm{neg}}$} & {$\chi$} \\
\midrule
3 & -3.44e-5 & 1.15e-5 \\
5 & -1.90e-6 & 3.80e-7 \\
6 & -2.43e-6 & 4.05e-7 \\
\bottomrule
\end{tabular}
\caption{Total negative baryon charge $B_{\mathrm{neg}}$ and its fraction $\chi=|B_{\mathrm{neg}}|/B$ of the total baryon number, at $H=0.00$.}
\label{tab:neg_charge}
\end{table}

\section{Conclusions}\label{sec:conclusions}

We have studied static solutions of the $U(1)$ gauged Skyrme-Maxwell model in a constant external magnetic induction $H\hat{z}$, for massless skyrmions with baryon number $B\leq 6$ and $H\in[0,8]$. These solutions were obtained with a GPU implementation of the Arrested Newton Flow on a $280^3$ lattice with spacing $0.1$ and fourth-order finite differences.

The energy of the gauged skyrmions is lower than in the ungauged model. For every value of $H$ the energy per baryon number decreases with growing $B$, so the binding energy relative to $B=1$ increases in the presence of the field. The energy curves show a global minimum at a field $H_{min}$ that also grows with $B$. This minimum is not interpreted from a topological constraint but as a balance in the Lagrangian terms which scale with different powers of $H$. The deviation of our solutions from the theoretical virial equilibrium \cref{derrick_energy} stays below $0.3\%$ of the energy per baryon number.

In the $B=1,2$ sectors, the solution is a smooth deformation of the ungauged rational-map configuration, while the symmetry axis is rotated to align with the field direction $\hat{z}$. A weak external field makes the skyrmion bigger than the ungauged one and a strong field makes it smaller. The volume peaks at $H\simeq 0.4$. The solutions also become more prolate along $\hat{z}$ as $H$ increases. This resembles the deformation produced by an anisotropic potential that breaks the isospin symmetry, a role played here by the magnetic field.

For $B>2$ and weak applied field, the geometry of the skyrmion is also a smooth deformation of a ungauged rational-map shape. However, over a narrow range of intermediate values surrounding $H=0.15$, we find that solutions form straight chains that are piled perpendicular to $\hat{z}$. For even higher magnetic fields these chains degenerate smoothly to toroidal shapes with their symmetry axis aligned in $\hat{z}$. In the $B\!=\!4$ case, other geometries compete in energy with the rational-map cube and the chain. 

The solutions with $B\!=\!3,5,6$ that are small rational-map deformations keep the folded structures of the ungauged fields, giving rise to regions of negative baryon density that form a core at the origin with tubular branches extending outward. No negative baryon density has been found in other geometries different than these.

The evaluation of the solutions provides a natural setting to test nuclear matter in the presence of high external magnetic fields. Recovering the SI units,

\begin{equation}
    H_{\mathrm{SI}} = \dfrac{a_0^2 f_\pi^2}{4e \hbar c^2} H \simeq 1.6479 \cdot 10^{15}\, H.
\end{equation}

Deformation becomes appreciable for $H_{\mathrm{SI}}\sim\SI{e14}{\tesla}$ and in this region various energy minimizing geometries emerge. These fields exceed those estimated at magnetar surfaces $\sim\SI{e11}{\tesla}$, but they could be realized in their interiors, or already in heavy-ion collisions which reach $\sim\SI{e14}{\tesla}$. The model's predictions are therefore still beyond direct experimental reach. Nevertheless, they demonstrate that strong electromagnetic backgrounds can reshape the geometry of nuclei.

\section*{Acknowledgements}
CA and ARA acknowledge financial support from the Spanish Research State Agency (AEI) under project PID2023-152762NB-I00, from the Xunta de Galicia under the project ED431F 2023/10 and the CIGUS Network of Research Centres, from the María de Maeztu grant CEX2023-001318-M funded by MICIU/AEI /10.13039/501100011033, and from the European Union ERDF. ARA is further supported by predoctoral grant ED481A from the Consellería de Cultura, Educación, Formación Profesional e Universidades, Xunta de Galicia. ARA thanks the University of Edinburgh, and in particular the School of Mathematics, for their hospitality during a research stay, where part of this work was carried out.

\appendix
\renewcommand{\thesection}{\Alph{section}}
\titleformat{\section}
  {\centering\normalsize\bfseries}
  {APPENDIX \thesection.}
  {0.5em}
  {}

\section{Visual representation}\label{sec:visual_rep}

All the shown 3D plots in this work represent isosurfaces of constant baryon density $B^0_{iso}$. This threshold value is chosen such that the isosurface encloses the 90\% of the total baryon charge, surrounding all lattice points with $B^0 \geq B^0_{iso}$.

\begin{figure}[h!]
  \centering
  \includegraphics[width=0.8\linewidth]{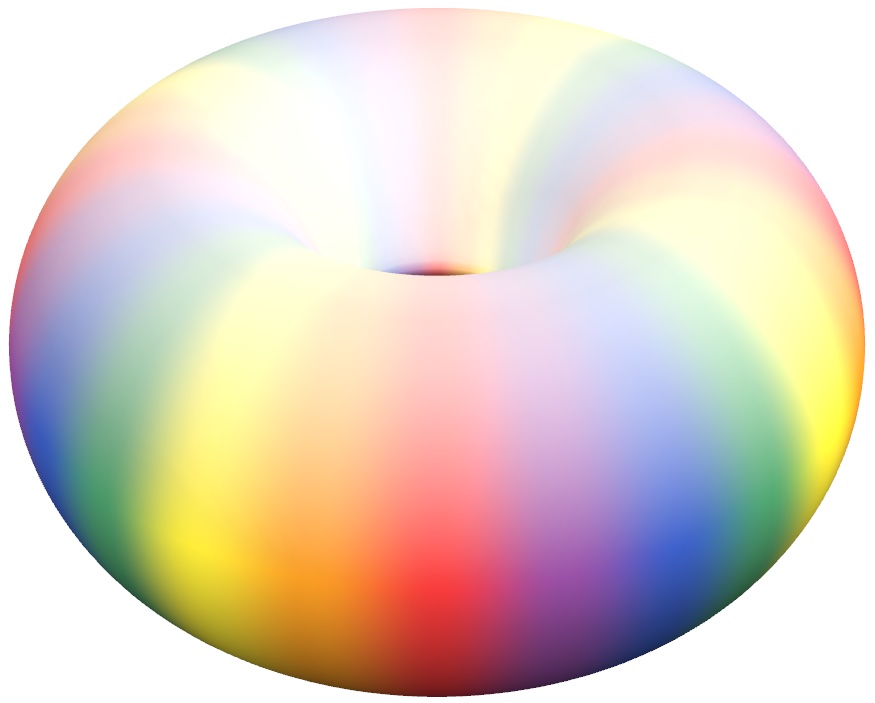}
  \caption{Topological density isosurface for $B=4$ and $H = 0.5$.}
  \label{fig:visual_repr}
\end{figure}

The isosurfaces are coloured according to the Runge's sphere, which allows us to visually distinguish the winding of the solutions. For instance in \cref{fig:visual_repr} we can check that each colour stripe repeats four times, confirming $B=4$. In this example the stripes warp around the smaller circles of the toroid, rather than around the longitudinal circle.

The colouring is defined in terms of the normalized pion triplet $(\phi_1,\phi_2,\phi_3)/\sqrt{\phi_i \phi_i}$, mapped to a custom HSL colour space. The hue is set by the azimuthal angle in the $(\phi_1,\phi_2)$ plane,
\begin{equation}
\text{hue} = \dfrac{\operatorname{atan2}(\phi_2,\phi_1) + \pi}{2\pi},
\end{equation}
while the lightness depends only on the $\phi_3$ field,
\begin{equation}
\text{lightness} = 0.90 \left(\dfrac{1-\phi_3}{2}\right).
\end{equation}

The node colours obtained on the equator $\phi_3 = 0$, together with the almost black and white poles at $\phi_3 \pm 1$, are listed in \cref{tab:runge_colors}.

\begin{table}[h!]
\centering
\begin{tabular}{lc}
\hline
Colour & $(\phi_1, \phi_2, \phi_3)/\sqrt{\phi_i \phi_i}$ \\
\hline
White  & $(0,\ 0,\ -1)$ \\
Green  & $(1,\ 0,\ 0)$ \\
Yellow & $\frac{1}{2}(1,\ -\sqrt{3},\ 0)$ \\
Orange & $\frac{1}{2}(-1,\ -\sqrt{3},\ 0)$ \\
Red    & $(-1,\ 0,\ 0)$ \\
Violet & $\frac{1}{2}(-1,\ \sqrt{3},\ 0)$ \\
Blue   & $\frac{1}{2}(1,\ \sqrt{3},\ 0)$ \\
Black  & $(0,\ 0,\ 1)$ \\
\hline
\end{tabular}
\caption{Correspondence between field values and the node colors of the Runge sphere.}
\label{tab:runge_colors}
\end{table}

\printbibliography

\end{document}